%% file: main.tex
\documentclass[journal]{IEEEtran}
\usepackage[noadjust]{cite}

\ifCLASSINFOpdf
  \usepackage[pdftex]{graphicx}
\else
  \usepackage[dvips]{graphicx}
\fi

\usepackage{amsmath}
\usepackage{algorithmic}
\usepackage{array}
\ifCLASSOPTIONcompsoc
 \usepackage[caption=false,font=normalsize,labelfont=sf,textfont=sf]{subfig}
\else
 \usepackage[caption=false,font=footnotesize]{subfig}
\fi

\usepackage{multirow}
\usepackage{amsfonts}
\usepackage[colorlinks=true,linkcolor=blue,citecolor=blue,anchorcolor=blue]{hyperref}
\usepackage{url}
\usepackage{booktabs}
\usepackage{color}
\usepackage{xcolor}
\usepackage[utf8]{inputenc}
\usepackage{cleveref}

\newcommand{\etal}{\textit{et al.}}

\begin{document}

%
\title{Beyond the Status Quo: A Contemporary Survey of Advances and Challenges in Audio Captioning}
%
%
%

\author{Xuenan Xu,~\IEEEmembership{Student Member,~IEEE},
        Zeyu Xie,
        Mengyue~Wu$\dagger$,~\IEEEmembership{Member,~IEEE},
        Kai Yu$\dagger$,~\IEEEmembership{Senior Member,~IEEE\thanks{The authors are from X-LANCE Lab, Department of Computer Science and Engineering, MoE Key Lab of Artificial Intelligence, AI Institute, Shanghai Jiao Tong University. $\dagger$Mengyue Wu and Kai Yu are the corresponding authors.}}
\thanks{}}

\markboth{}
{Shell \MakeLowercase{\textit{et al.}}: Bare Demo of IEEEtran.cls for IEEE Journals}



\maketitle

\begin{abstract}
Automated audio captioning (AAC), a task that mimics human perception as well as innovatively links audio processing and natural language processing, has overseen much progress over the last few years. 
AAC requires recognizing contents such as the environment, sound events and the temporal relationships between sound events and describing these elements with a fluent sentence.
Currently, an encoder-decoder-based deep learning framework is the standard approach to tackle this problem.
Plenty of works have proposed novel network architectures and training schemes, including extra guidance, reinforcement learning, audio-text self-supervised learning and diverse or controllable captioning.
Effective data augmentation techniques, especially based on large language models are explored. 
Benchmark datasets and AAC-oriented evaluation metrics also accelerate the improvement of this field.
This paper situates itself as a comprehensive survey covering the comparison between AAC and its related tasks, the existing deep learning techniques, datasets, and the evaluation metrics in AAC, with insights provided to guide potential future research directions.
\end{abstract}


%
\IEEEpeerreviewmaketitle

\input{chapters/intro}

\input{chapters/related_tasks}

\input{chapters/encoder_decoder}

\input{chapters/training_schemes}

\input{chapters/data_augmentation}

\input{chapters/datasets_evaluation}

\input{chapters/future_direction}

\input{chapters/conclusion}


%



\section*{Acknowledgment}

This work has been supported by National Natural Science Foundation of China (No.61901265), Shanghai Municipal Science and Technology Major Project (2021SHZDZX0102), Key Research and Development Program of Jiangsu Province, China (No.BE2022059-2).

\ifCLASSOPTIONcaptionsoff
  \newpage
\fi



\bibliographystyle{IEEEtran}
\bibliography{refs}

%

\begin{IEEEbiography}[{\includegraphics[width=1in,height=1.25in,clip,keepaspectratio]{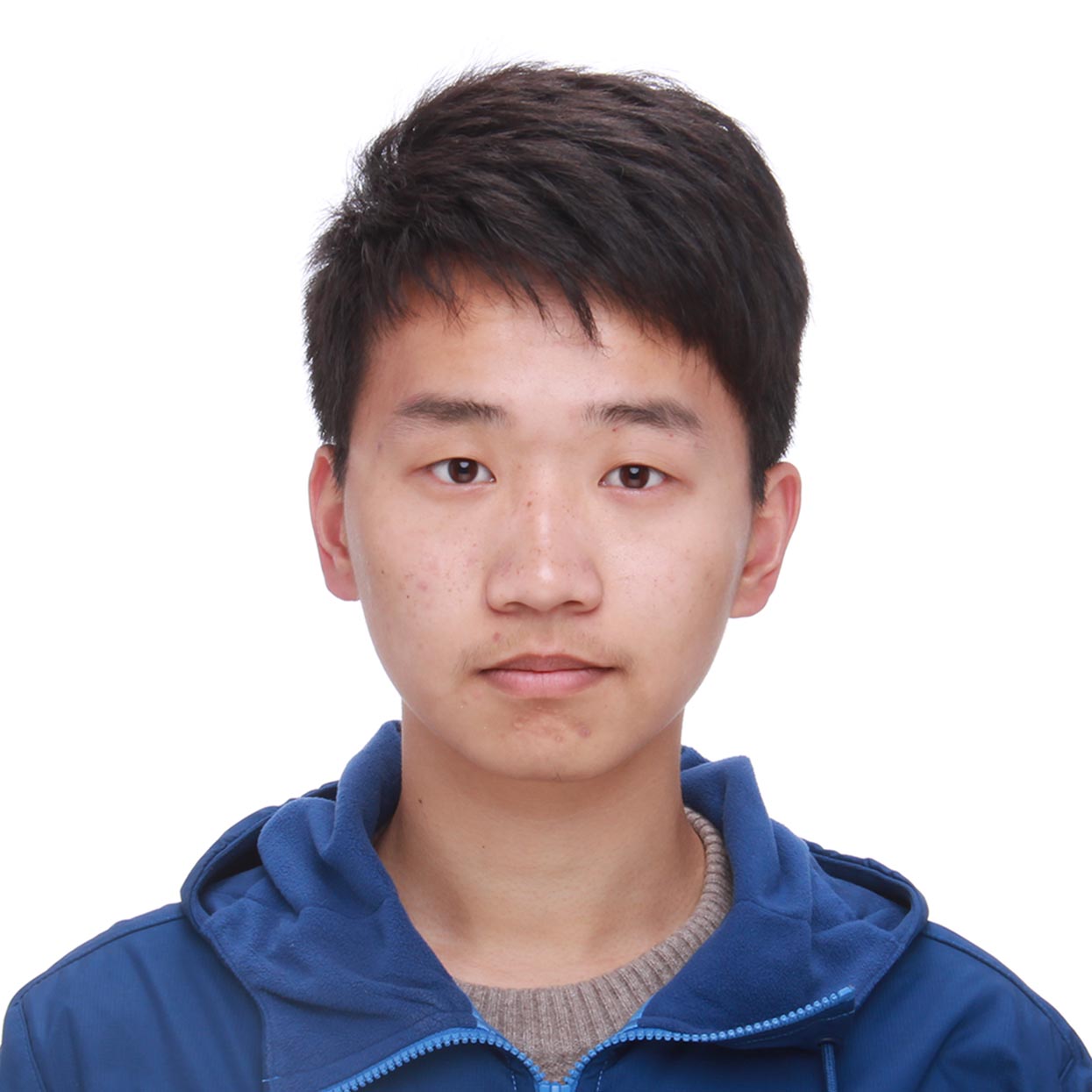}}]{Xuenan Xu}
received his B.S. degree from Shanghai Jiao Tong University in 2019. He is currently working towards his Ph.D. degree in Shanghai Jiao Tong University. His supervisors are Kai Yu and Mengyue Wu. His main research interests include detection and classification of acoustic scenes and events, the interaction between audio signal processing and natural language processing, multimodal interaction between audio, language and vision.

\end{IEEEbiography}

\begin{IEEEbiography}[{\includegraphics[width=1in,height=1.25in,clip,keepaspectratio]{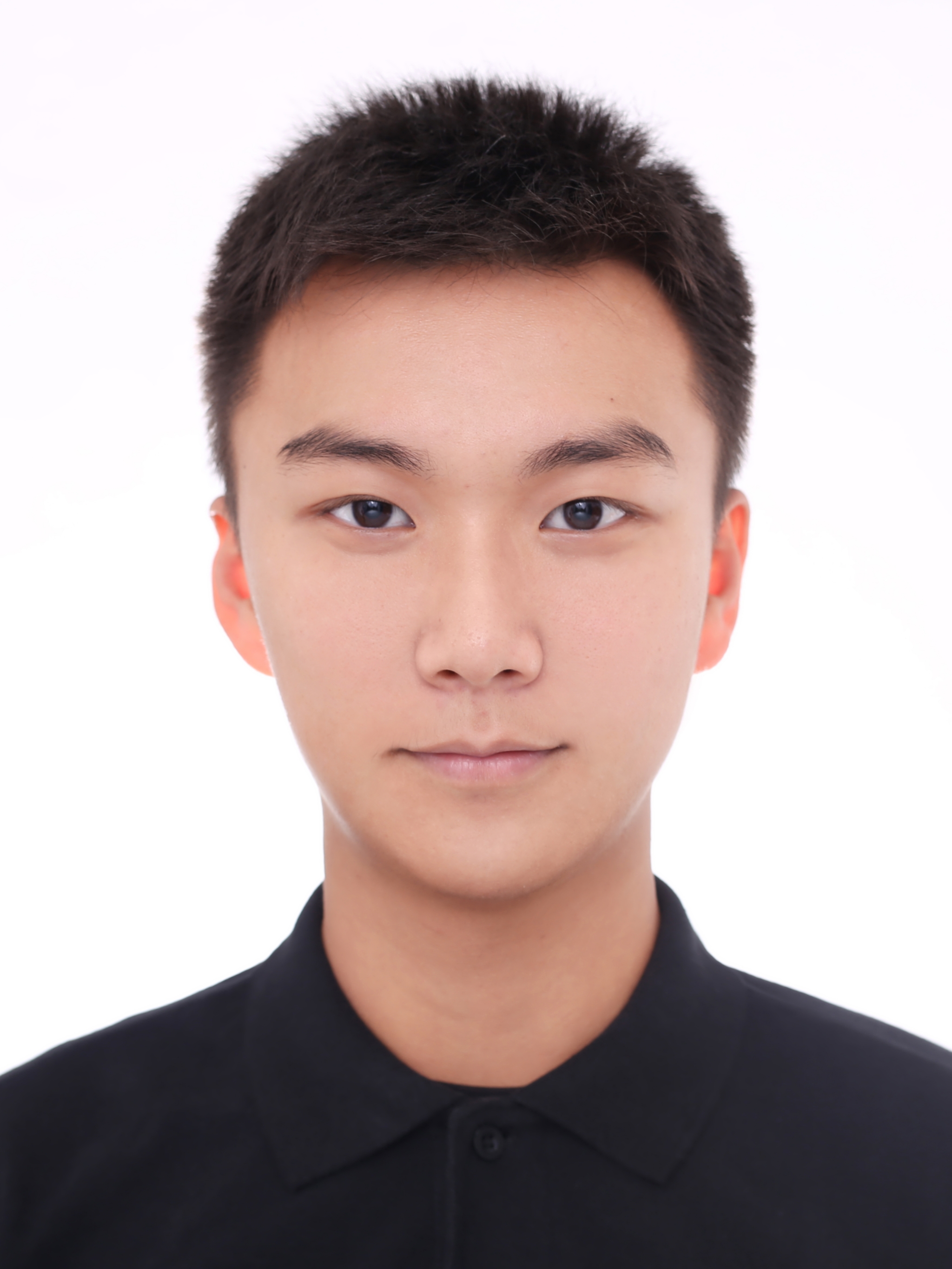}}]{Zeyu Xie}
received the B.S. degree in 2022 from Shanghai Jiao Tong University, Shanghai, China, where he is currently working toward the M.Sc. degree. His supervisors are Kai Yu, and Mengyue Wu. His main research interests include the interaction between  audio signal processing and natural language processing, foley sound synthesis guided by multi-modal information.

\end{IEEEbiography}

\begin{IEEEbiography}[{\includegraphics[width=1in,height=1.25in,clip,keepaspectratio]{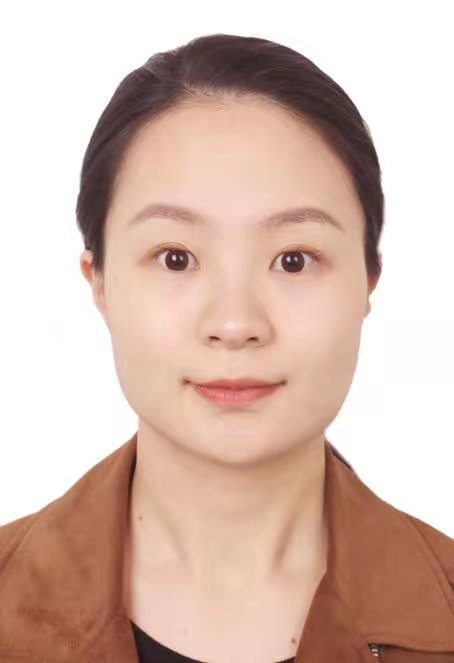}}]{Mengyue Wu}
received her B.S. and B.A. from Beijing Normal University in 2011 and was awarded Ph.D. from the University of Melbourne in 2017. She is currently an Assistant Professor in Computer Science and Engineering Department, Shanghai Jiao Tong University, China. Her main research interests lie in the area of audio- and language- based human machine interaction including audio processing, multimedia processing, and medical application of these technologies. 
\end{IEEEbiography}

\begin{IEEEbiography}[{\includegraphics[width=1in,height=1.25in,clip,keepaspectratio]{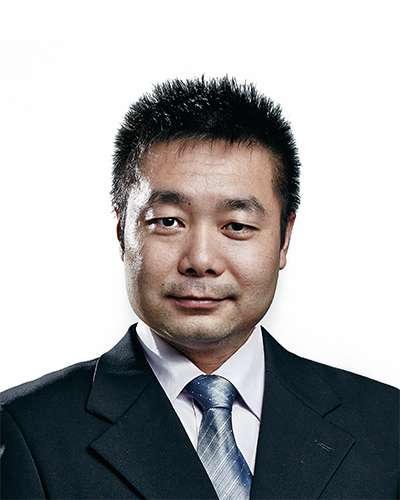}}]{Kai Yu}
is a professor at Computer Science and Engineering Department, Shanghai Jiao Tong University, China. He received his B.Eng. and M.Sc. from Tsinghua University, China in 1999 and 2002, respectively. He then joined the Machine Intelligence Lab at the Engineering Department at Cambridge University, U.K., where he obtained his Ph.D. degree in 2006. His main research interests lie in the area of speech-based human machine interaction including speech recognition, synthesis, language understanding and dialogue management. He is a member of the IEEE Speech and Language Processing Technical Committee.
\end{IEEEbiography}

\end{document}

%% file: chapters/intro.tex
\section{Introduction}

\label{sec:intro}
\IEEEPARstart{B}{eing} one of the five senses (i.e., sight, hearing, touch, smell, and taste), auditory information provides us with rich knowledge.
We obtain critical information to interact with the physical world via sounds around us.
A machine with a human-like auditory perception system should be able to comprehensively process auditory signals, including speech and non-speech sounds (e.g., music, animal sounds, mechanical sounds).
Recent audio research oversees a gradually increasing amount of attention on general non-speech sounds, a critical part of the auditory information processed by the human ear. 
Compared with speech-focused tasks (e.g., speech recognition and synthesis), such non-speech research is still at an emerging stage. 
Detection and Classification of Acoustic Scenes and Events (DCASE) challenges have greatly encouraged research improvements in this field by providing benchmark datasets and adding in new tasks. 
Unlike speech, which has a relatively limited phonetic system, non-speech sounds involve a much larger category. 
This has certainly led to difficulty and complexity in task definitions.
For example, an audio can be described from different aspects, e.g., sound events, relationships between sound events, and acoustic environments.

\begin{figure*}[!htpb]
    \centering
    \includegraphics[width=\textwidth]{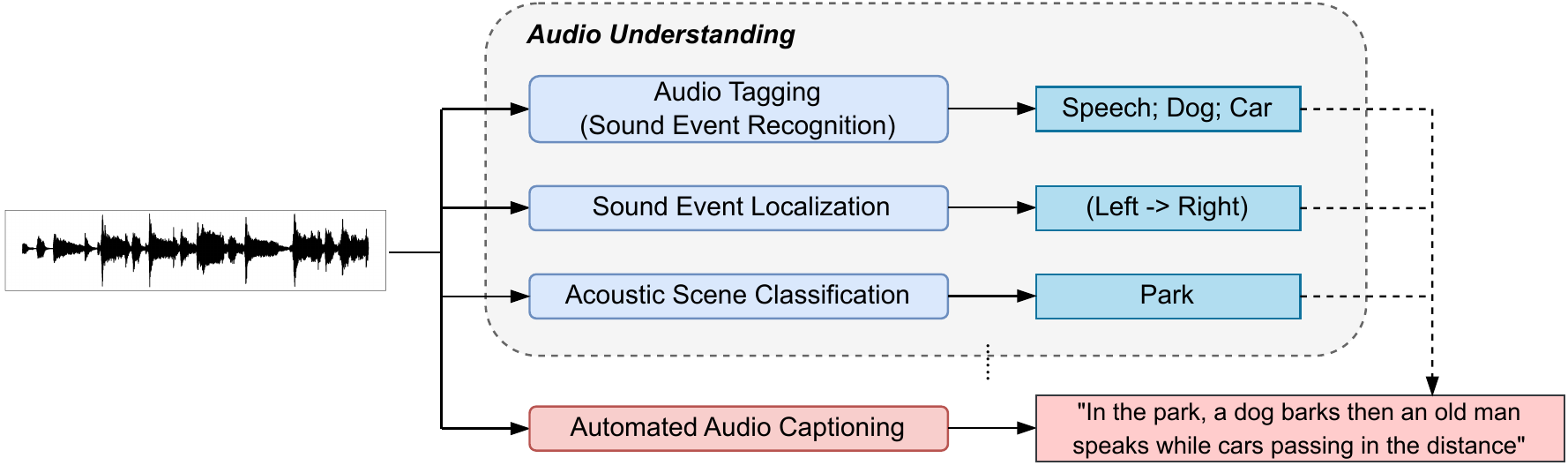}
    \caption{An overview of audio understanding tasks, including low-level sound event recognition and localization to high-level scene / environment recognition. Automated audio captioning is a text generation task based on these audio understanding tasks.}
    \label{fig:audio_understanding_tasks}
\end{figure*}
Compared with class labels, natural language description conveys unrestricted and richer information, as shown in \Cref{fig:audio_understanding_tasks}, which is more like human perception.
Therefore, such a task called automated audio captioning (AAC) is proposed, with the purpose of generating a textual description (caption) for the given audio signal input. 
Unlike speech processing tasks where non-speech sound events are taken as noise, AAC requires analyzing and describing all sound events.
The caption is a summarization of the audio content.
Through the analysis and statistical examination of captions in current AAC datasets, we summarize elements in captions and their distribution as follows:
\begin{itemize}
    \item low-level sound event attributes, e.g., a \textbf{loud} barking, a gun is shot \textbf{multiple times}, a \textbf{high-pitched} engine running
    \item sound event types, e.g., \textbf{dog barking} and \textbf{a man speaks}
    \item relationships between sound events, e.g., event A \textbf{followed by} event B
    \item induced human emotion, e.g., an \textbf{annoying} fly
    \item high-level description or inference of the environment, e.g., a man speaking \textbf{in the room} 
    \item event-specific details, e.g., speech contents, music styles 
\end{itemize}
The above elements are listed in ascending order, from low-level to high-level.
They follow a long-tailed distribution: low-level elements are frequent in captions while high-level elements are highly uncommon.
A typical example of an audio caption is ``\texttt{outdoor a man says yes loudly followed by an annoying fly buzzing in the background.}''

Compared with classic audio understanding tasks like acoustic scene classification and audio tagging, AAC is relatively new and rarely investigated until recently.
In 2019 several AAC datasets were published~\cite{wu2019audio,kim2019audiocaps,lipping2019crowdsourcing}.
From 2020 to 2023, AAC has been incorporated as a task in DCASE challenges to encourage more studies in exploring novel model architectures and training schemes.

AAC research has benefited from its closely-related tasks such as audio understanding and cross-modal text generation tasks, which will be discussed in \Cref{sec:related_tasks}.
Compared with visual captioning, sounds carry complementary yet rich information that is unavailable via visual analysis.
For example, a truck can be recognized in a video but the audio is required to determine whether it honks.
Audio signals are more reliable for perception when visual signals are noisy (e.g., in dark environments, objects are occluded). 
Compared with audio understanding tasks, AAC generates unrestricted natural language, a more suitable medium for human-machine interaction.
Therefore, many applications can be derived from AAC, such as automatic description generation for Internet videos (e.g., Youtube, Flickr), interpretable surveillance or monitoring systems, and audio retrieval based on natural language query~\cite{oncescu2021audio}.

Although AAC research has attracted a certain amount of attention \footnote{A list of related papers and tools can be found in \url{https://github.com/audio-captioning/audio-captioning-papers} and \url{https://github.com/audio-captioning/audio-captioning-resources}}, a comprehensive survey on its definition, development, challenges and future directions is still lacking.
A contemporary work~\cite{mei2022automated} reviews fundamental topics of AAC, such as model architectures, auxiliary information and datasets\footnote{Works until April 2022 were included in \cite{mei2022automated}}.
Deviating from mere summarization of different components and techniques, our review situates itself as a critical piece.
We delve into more profound topics from different aspects: why an approach can or cannot achieve good performance and why current models are not good enough, with deep insights.
Based on the new trends summarized from recent works such as utilizing large language models (LLMs) and modifications on evaluation metrics, we give our insights on the effectiveness of existing approaches and challenges of AAC. 
For fundamental topics, we refer readers to \cite{mei2022automated}.

Specifically, we state the main differences between this paper and \cite{mei2022automated}:
\begin{enumerate}
    \item We make a comparison between AAC and similar or related tasks in \Cref{sec:related_tasks}. The comparison between AAC and visual captioning highlights current AAC challenges.
    \item In \Cref{sec:encoder_decoder} and \Cref{sec:training_schemes}, we provide a succinct description of contents covered in \cite{mei2022automated}, including encoder and decoder architectures, extra guidance and reinforcement learning. The emphasis lies on new techniques, such as audio-text self-supervised learning.
    \item In \Cref{sec:data_augmentation}, we summarize data augmentation approaches, especially methods based on LLMs.
    \item In \Cref{sec:datasets_metrics}, we comprehensively compare current datasets, regarding the audio source and caption style. In terms of caption diversity, we provide a picture that differs from \cite{mei2022automated}.
    \item In \Cref{sec:datasets_metrics}, several recently-proposed metrics for AAC are introduced and their respective application scenarios are summarized.
    \item In \Cref{sec:future_directions}, we elaborate on several potential directions according to the limitations of current AAC approaches, which we believe are crucial challenges to be addressed for the application of AAC.
\end{enumerate}

Hence, this paper aims to fill the gap by providing a detailed survey of AAC, with insights from comparison with similar tasks, existing approaches, datasets, evaluation metrics and future directions.

The remainder of this paper is organized as follows:
\Cref{sec:related_tasks} introduces audio understanding and cross-modal text generation tasks since they are closely related to AAC.
\Cref{sec:encoder_decoder} presents the mainstream encoder-decoder AAC architecture.
\Cref{sec:training_schemes} presents a detailed discussion of training schemes and topics involved in existing works.
These works explored training schemes or objectives different from the standard encoder-decoder training recipe.
\Cref{sec:data_augmentation} gives a review of data augmentation methods in AAC.
\Cref{sec:datasets_metrics} presents public datasets, commonly used evaluation metrics, and a comparison of them.
Finally, a brief discussion on future research directions is given in \Cref{sec:future_directions} and \Cref{sec:conclusion} ends the review with a conclusion.

%% file: chapters/related_tasks.tex
\section{Related Tasks}
\label{sec:related_tasks}
AAC naturally involves two sub-tasks: audio understanding and text generation.
Here audio understanding refers to audio tasks that do not involve natural language, such as classification and localization.
Both tasks had been extensively explored before the proposal of AAC, thus methods employed in these tasks serve as valuable inspiration for AAC.
Moreover, comparing AAC with its related tasks enables us to better understand its specific characteristics and challenges.
Therefore, in this section, we present tasks related to AAC, providing a comparison in terms of objectives, data, and methods.
Audio understanding involves recognizing sound event types and properties, acoustic scenes and so on, which is the first step for summarizing the audio.
Since the second step of AAC is text generation conditioned on audio embeddings, we present another cross-modal text generation task: visual captioning.

\subsection{Audio Understanding}
\label{subsec:audio_understanding_tasks}
Audio understanding involves a wide range of tasks, among which some AAC-relevant ones are listed below:
\begin{itemize}
    \item Audio tagging (AT): recognizing sound events that occurred in an audio clip, where multiple sound events can co-occur
    \item Sound event detection (SED): predicting event time boundaries of each detected event in addition to AT
    \item Sound event localization (SEL): predicting the spatial trajectories of each detected event in addition to sound event detection
    \item Sound emotion recognition (SER): recognizing emotional states conveyed or induced by sound events
    \item Acoustic scene classification (ASC): recognizing the surrounding environments
\end{itemize}
Although these tasks focus on different objectives, they often share similar audio encoding backbones, which are taken as audio encoders for AAC (see \Cref{subsec:audio_encoder}).
Moreover, some works utilize the output from these audio understanding tasks as an extra input to caption generation for guidance (discussed in \Cref{subsec:extra_guidance}).
To this end, we give a brief introduction of common backbones and datasets for audio understanding tasks.

Despite early exploration of non-neural network approaches, convolutional neural networks (CNN) and their combination with the recurrent neural networks (RNN), called CRNN, remained the state-of-the-art (SOTA) framework for a long period.
For classification tasks like AT and ASC, CNN achieves satisfactory results while CRNN is suitable for SED due to the temporal modelling ability of RNN.
With the success of vision Transformers~\cite{dosovitskiy2020image}, there is a trend of applying Transformer to audio understanding tasks~\cite{gong2021ast,koutini2022efficient} since Transformer is suitable for both classification and time dependency modelling.
Source separation, where both classical statistical approaches~\cite{wang2006computational} and neural networks~\cite{wisdom2020unsupervised,petermann2023tackling,li2023target} can help, is also extensively explored to separate sound event mixtures for subsequent processing.
However, the attempt in SED indicates that source separation provides little help in audio understanding~\cite{liang2022selective,huang2020guided}.
The mainstream approach to audio understanding tasks does not incorporate source separation as a pre-processing step.

A milestone dataset for audio understanding is the release of AudioSet~\cite{gemmeke2017audio}, a large-scale weakly-annotated sound event dataset.
Since it is by far the most large-scale annotated public audio dataset, many works pre-train an audio encoding backbone in a supervised or self-supervised way before transferring to the target task~\cite{kong2020panns,wu2023beats}.

\subsection{Cross-modal Text Generation}
\label{subsec:visual_captioning}

\begin{figure*}[ht]
    \centering
    \includegraphics[width=0.8\linewidth]{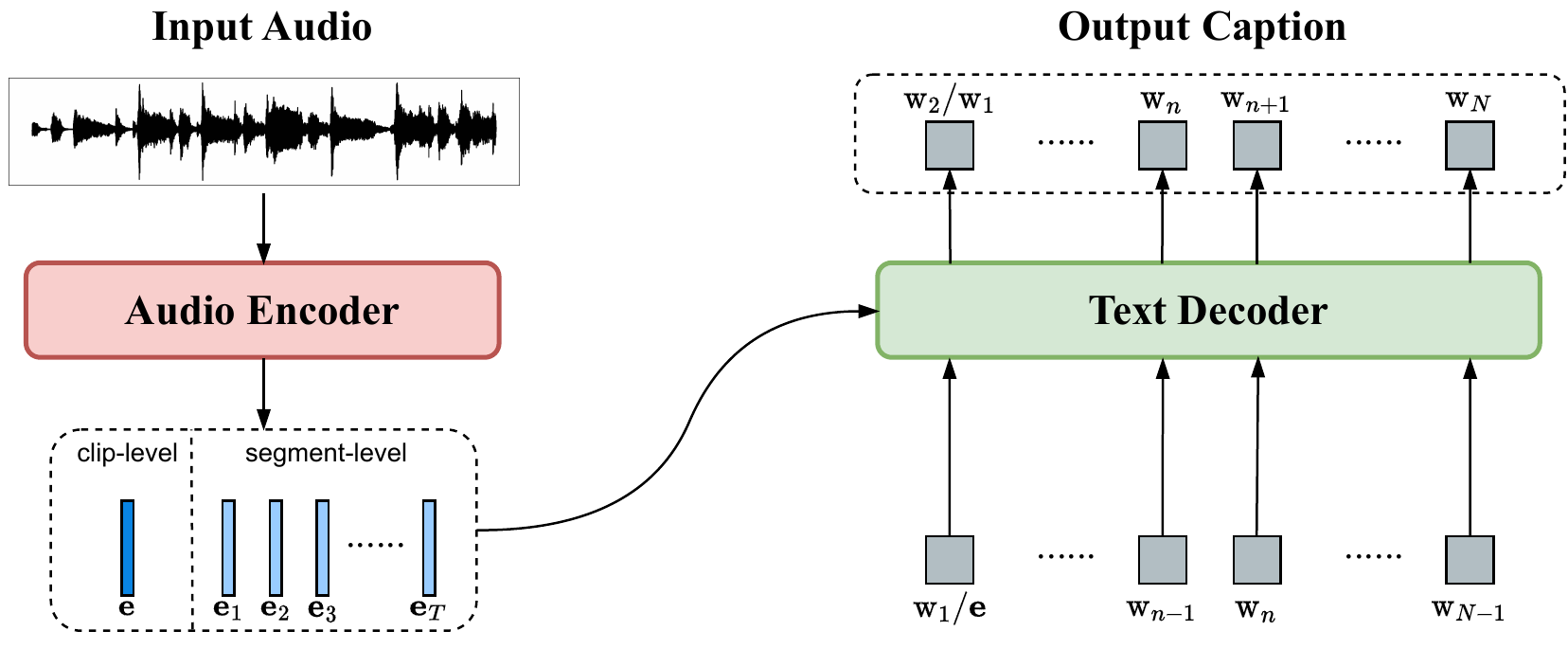}
    \caption{The illustration of encoder-decoder AAC architecture. The encoder transforms the input audio feature to a clip-level embedding $\mathbf{e}$ or segment-level embedding sequence $\{\mathbf{e}_t\}_{t=1}^T$. The decoder takes the partly generated words and audio embeddings to predict the word of this timestep.}
    \label{fig:encoder_decoder}
\end{figure*}

In this part, we discuss analogous cross-modal text generation in computer vision: visual captioning, including image and video captioning.
They are very similar to AAC with the only difference in the input (audio vs. visual signal).
Compared with AAC, visual captioning has been explored since the early days of deep learning.
Before the wide application of neural networks (NNs), researchers proposed template-based~\cite{hodosh2013framing} and retrieval-based~\cite{farhadi2010every} methods for image captioning.
With the success of deep learning, the NN-based encoder-decoder framework has been the standard practice for visual captioning.
Similar to AAC, visual captioning also aims to generate text conditioned on information from another modality.
Therefore, AAC text decoders (RNN and Transformer, discussed in \Cref{subsec:text_decoder}) are mainly adapted from visual captioning.
In addition, some training schemes (e.g., reinforcement learning, diverse captioning) and evaluation metrics of AAC are also derived from visual captioning (see \Cref{sec:training_schemes} and \Cref{subsec:eval_metrics}).
As text generation tasks, both audio and visual captioning suffer from unreliable automatic evaluation metrics (see \Cref{subsec:eval_metrics}).
Visual and audio captioning also interact in the modality fusion in the video.
Some video captioning works use audio as auxiliary information~\cite{tian2018attempt,xu2017learning} while the inclusion of visual clues to guide audio captioning is also explored~\cite{liu2022visually}.

However, there are significant differences between audio and visual captioning.
The differences are reflected in datasets.

\begin{itemize}
    \item First, in terms of the dataset size, AAC datasets are much smaller than visual captioning data.
    The most large-scale AAC dataset AudioCaps contains about 51K audio clips and 57K caption annotations (see \Cref{tab:datasets}).
    In comparison, the image captioning dataset Microsoft Common Objects in COntext (MSCOCO)~\cite{lin2014microsoft} contains more than 300K images.
    The video captioning dataset MSR-Video to Text (MSR-VTT) contains fewer video clips (10K) but much more caption annotations (200K).
    Moreover, there are naturally many more image-text or video-text pairs available on the web than audio-text pairs.
    \item Second, visual and audio captions can be both diverse but the two types of diversity are caused by different factors.
    Visual captions are diverse in the objects or events chosen to describe.
    There are often too many objects (events) so that the single-sentence caption can only cover part of them.
    Different captions of the same image or video are all correct but they do not cover all the visible elements.
    In contrast, the diversity of audio captions lies in the ambiguity of sounds and the inference of listeners.
    Different sources can produce similar sounds (i.e., paper and leaf can both make rustling sounds) so listeners can only infer the objects from an audio.
    As a result, multiple captions of the same audio clip are diverse but they do not necessarily all accurately describe the actual sound event.
    Without the aid of visual information, the actual event may be recognized as an acoustically similar one. 
    \item Third, details in image and video are much more discernable than those in audio so visual captions are often more detailed than audio captions.
    For example, while annotators seldom neglect the relationships between objects in images and videos (e.g., using ``a girl and a chair'' instead of ``a girl sitting on a chair''), they tend to ignore the temporal relationships between sound events in audios.
    Statistics show that only 24.5\% of captions in AudioCaps involve specific temporal expressions.
    In addition to the issue of annotation instructions that do not prompt annotators to describe temporal relationships, the fact that visual details are more perceptible than audio details and listening to an audio is more time-consuming than observing an image may also lead to this gap.
\end{itemize}
Due to these differences, successful approaches in visual captioning may not be directly adapted to AAC.
Specifically, the second and the third differences listed above often receive less attention from researchers.
These differences pose respective challenges to AAC: data scarcity, the lack of diversity and details in captions.
Researchers propose several data augmentation approaches to tackle the data scarcity problem, which will be presented in \Cref{sec:data_augmentation}.
However, current methods and datasets are not sufficient to achieve diverse and detailed captioning.
\Cref{subsec:future_controllable_captioning} and \Cref{subsec:future_detailed_captioning} further elaborate on this potential direction.

%% file: chapters/encoder_decoder.tex
\section{Encoder-decoder Based AAC}
\label{sec:encoder_decoder}
Following the NN-based standard practice mentioned in \Cref{subsec:visual_captioning}, current audio captioning systems all adopt the sequence-to-sequence encoder-decoder framework.
In this section, we give a general overview of the encoder-decoder framework and the network architectures\footnote{For more detailed introductions on network architectures, we refer readers to \cite{mei2022automated}.}.
Then, we give a detailed introduction of transferring pre-trained models.

\subsection{Encoder-decoder Framework Overview}
In the encoder-decoder framework, the two sub-tasks of AAC, audio understanding and text generation, are accomplished by the audio encoder and the text decoder, respectively.
\Cref{fig:encoder_decoder} illustrates the standard encoder-decoder approach.
For an input audio $\mathcal{A}$, the encoder transforms it into a clip-level embedding $\mathbf{e}$ and a segment-level embedding sequence $\{\mathbf{e}_t\}_{t=1}^T$.
The decoder auto-regressively generates the caption $\{w_i\}_{i=1}^{N}$, consisted by $N$ words, based on audio embeddings.
The whole model is trained end-to-end by cross-entropy (XE) loss:
\begin{align}
    \mathbf{e},\; &\{\mathbf{e}_t\}_{t=1}^T = \text{Enc}(\mathcal{A})\\
    p_n &= \text{Dec}(\{w_i\}_{i=1}^{n-1},\; \mathbf{e},\; \{\mathbf{e}_t\}_{t=1}^T)\\
    \mathcal{L} &= -\sum_{n=1}^N \log p_n(\hat{w}_n)
\end{align}
where $p_n(\hat{w}_n)$ is the predicted probability of the ground truth word $\hat{w}_n$.
Some works incorporate additional loss items to facilitate training, which will be discussed in \Cref{subsec:extra_guidance}.

\subsection{Audio Encoder}
\label{subsec:audio_encoder}

The audio encoder extracts embeddings from the input audio for caption generation.
It usually consists of a non-NN-based feature extractor followed by a neural network.
We first briefly introduce the hand-crafted features, and then discuss the neural network architectures.

The extraction of hand-crafted features is crucial for signal processing tasks, especially before the era of deep learning.
Here we list features explored in AAC.
\begin{itemize}
    \item log mel-spectrogram (LMS): the most widely-used feature in AAC, extracted by applying mel-scale filterbanks followed by a logarithmic transformation.
    \item mel-frequency cepstral coefficients (MFCC): typically used in speech recognition, extracted by applying a discrete cosine transform (DCT) to the mel-spectrogram.
    \item Gammatone filter bank: designed to model the human auditory perception, by applying Gammatone filterbanks instead of mel-scale filterbanks.
    \item harmonic percussive source separation (HPSS): typically used in music processing tasks, extracted by separating the harmonic and percussive components.
    \item Pitch and fundamental frequency (F0): typically used in speech and music processing, extracted via various estimation algorithms.
\end{itemize}
Among them, features except LMS are only utilized in early works under the situation of very limited data~\cite{ikawa2019neural,perez-castanos2020listen,koizumi2020ntt} as they do not present significant advantages over LMS.
Compared with features like HPSS, LMS is a relatively general feature for audio processing tasks so it fits AAC's purpose in recognizing various kinds of sound events and achieving a wide range of goals illustrated in \Cref{sec:intro}.

In the following part, mainstream audio encoder networks are presented, including RNN, CNN, CRNN and Transformer.

\paragraph{RNN}
RNNs are suitable for signal processing because of their ability to process variable-length sequences.
The two variants, long short-term memory (LSTM) networks and gated recurrent unit (GRU) networks, are commonly used.
The LMS feature with $\mathcal{T}$ frames is viewed as a sequence of length $\mathcal{T}$.
Early works use RNN to encode the sequence~\cite{drossos2017automated,wu2019audio,ikawa2019neural,eren2020semantic} but the performance falls far behind other encoders, possibly due to the over-large sequence length.

\paragraph{CNN}

CNNs are powerful in extracting position-invariant features~\cite{scherer2010evaluation} hence they perform well in image processing.
Since an audio spectrogram can be viewed as a one-channel image where sound events can be recognized from specific time-frequency patterns, CNNs are successfully adapted to many audio understanding tasks.
Therefore, CNNs remain one of the mainstream encoder architectures until now.
Some modifications are made to typical 2-dimensional (2D) CNNs:
\begin{itemize}
    \item 1D + 2D CNN: Researchers explored combining 1-dimension (1D) CNNs with 2D CNNs to incorporate temporal information into caption generation~\cite{tran2021wavetransformer,han2021automated}. 
    However, the combination did not bring significant improvement against 2D CNN encoders, indicating that 1D CNNs may not be suitable for temporal modelling.
    \item Feature augmenter: Some works augment the CNN-extracted features by multi-level fusion~\cite{kim2019audiocaps,ye2021improving} or appending a new module~\cite{xiao2023graph}. This practice often brings significant improvement over vanilla CNNs.
\end{itemize}
 
\paragraph{Transformer}
\label{subsubsec:encoder_transformer}
Transformers~\cite{vaswani2017attention} are the mainstream architecture for many natural language processing (NLP), due to their ability to model time dependencies in long sequences.
In AAC, vision Transformers (ViT) are adapted as audio encoders.
A spectrogram is transformed into a sequence of patches to be fed to ViT.
Transformer encoders achieved similar performance to CNN encoders~\cite{mei2021audio,kouzelis2022efficient}.
Analysis showed that pre-training is crucial for Transformer encoders to achieve good performance.

\paragraph{CRNN / CNN-Transformer}
Combining CNNs with RNNs or Transformers, CRNNs or CNN-Transformers encode both time-frequency and temporal information.
The embedding sequence output by CNNs is further encoded by RNNs or Transformers.
Compared with CNNs, CRNN and CNN-Transformers achieve comparable performance~\cite{xu2020crnn,koizumi2020ntt,koh2022automated}.
In some works, CNNs are combined with Transformer variants, like $M^2$ Transformer~\cite{chen2021m2transformer} and Conformer~\cite{narisetty2021leveraging}.
Their results showed that using pre-trained CNNs in CRNNs and CNN-Transformers is critical for the architecture to perform well.
When leveraging a pre-trained deep CNN, freezing its parameters while keeping RNN or Transformer trainable is a good practice (discussed in \Cref{subsubsec:encoder_or_decoder_pretraining}).

\subsection{Text Decoder}
\label{subsec:text_decoder}

The text decoder is a conditional language model to generate captions conditioned on the audio embeddings.
A word embedding layer transforms already generated words into fixed-size embeddings.
Weck \etal~\cite{weck2021evaluating} compared several pre-trained word embeddings and found that BERT embeddings perform the best.
However, results also showed that randomly initialized embeddings already achieve promising results.
Hereby, we focus on the network architecture after the word embedding layer and decoding strategies.
Since the caption is a word sequence, RNNs and Transformers are exclusively adopted as the architecture. 

\paragraph{RNN}
Uni-directional LSTM and GRU are commonly utilized as the text decoder~\cite{drossos2017automated,wu2019audio,ikawa2019neural}.
The attention mechanism is applied~\cite{xu2021investigating,ye2021improving} to make the decoder focus on different temporal parts of the audio at different timesteps.
Taking the previous ($\mathbf{h}_{n-1}$) or current hidden state ($\mathbf{h}_n$) as the \textit{query} and the audio embedding sequence $\{\mathbf{e}_t\}_{t=1}^T$ as both the \textit{key} and the \textit{value}, a context vector $\mathbf{c}$ is calculated:
\begin{equation}
    w_{t} = \frac{
    \mathrm{exp}\left(\mathrm{sim}(\mathbf{h}_{n-1}/\mathbf{h}_n, \mathbf{e}_t)\right)
    }
    {
    \sum_{t'=1}^{T}{\mathrm{exp}\left(\mathrm{sim}(\mathbf{h}_{n-1}/\mathbf{h}_n, \mathbf{e}_{t'})\right)}
    }, \mathbf{c} = \sum_{t=1}^{T}{w_{t}\mathbf{e}_{t}}
\end{equation}
Then $\mathbf{c}$ is fed to the decoder to predict the word probabilities.


\paragraph{Transformer}
Due to its popularity in NLP, Transformers have been replacing RNN as the decoder in recent works~\cite{chen2022icnn,mei2021audio,chen2022interactive}.
Transformers first use self-attention to encode word embeddings $\mathbf{e}_w$ into $\mathbf{h}_1$.
Then cross-attention is applied between $\mathbf{h}_1$ and audio embeddings $\mathbf{e}_a$.
Multi-head attention (MHA) is adopted in both steps.
\begin{align}
    \mathbf{h}_1 = \text{LayerNorm}(\mathbf{e}_w + \text{MHA}(\mathbf{e}_w, \mathbf{e}_w, \mathbf{e}_w))\\
    \mathbf{h}_2 = \text{LayerNorm}(\mathbf{h}_1 + \text{MHA}(\mathbf{h}_1, \mathbf{e}_a, \mathbf{e}_a))
\end{align}
Transformers in most AAC works are shallow, but some works also use deep ones ~\cite{chen2021m2transformer,gontier2021automated,narisetty2021leveraging} by incorporating large-scale datasets.
Xiao \etal~\cite{xiao2022local} proposed an attention-free Transformer to attend to both the local and global information.

\paragraph{
Decoding Strategies
}
Previous studies~\cite{takeuchi2020effects,wu2023beats} indicated that decoding or sampling strategies have a significant impact on captioning performance.
Takeuchi \etal~\cite{takeuchi2020effects} found that beam search decoding leads to significant improvement against greedy decoding.
Wu \etal~\cite{wu2023beats} found that nucleus sampling~\cite{holtzman2019curious} works better under the SPIDEr-FL metric (see \Cref{subsec:eval_metrics}).
The correlation between decoding strategies and evaluation metrics needs further exploration as decoding strategies directly influence the output caption.
Besides, reranking is also a technique to enhance performance.
Wu \etal~\cite{wu2023beats} combined the captioning model and a jointly-trained audio-text retrieval model to do reranking.
The former worked as a language model to give the likelihood while the latter provided the audio-text similarity.
Narisetty \etal~\cite{narisetty2021leveraging} implicitly reranked hypotheses by integrating a separate language model with the text decoder using shallow fusion~\cite{hori2017advances}.
In each decoding timestep, the decoder combines its score with the score provided by the language model to rerank current partly-decoded hypotheses.
Reranking commonly improves the captioning performance.
However, the improvement comes at the cost of significantly increased computational complexity and decreased decoding speed.

In summary, beam search decoding and nucleus sampling are commonly-used decoding strategies that strike a good balance between performance and decoding efficiency. 
The relatively marginal performance gains achieved by reranking make them impractical for real-world applications.

\subsection{Pre-training}
\label{subsubsec:encoder_or_decoder_pretraining}

Due to the data scarcity problem in AAC and the diversity of natural language, training the end-to-end AAC model from scratch produced unsatisfactory performance~\cite{xu2021investigating}.
Pre-training is now widely adopted to transfer powerful pre-trained uni-modal models to AAC.

\begin{figure}
    \centering
    \includegraphics[width=0.5\linewidth]{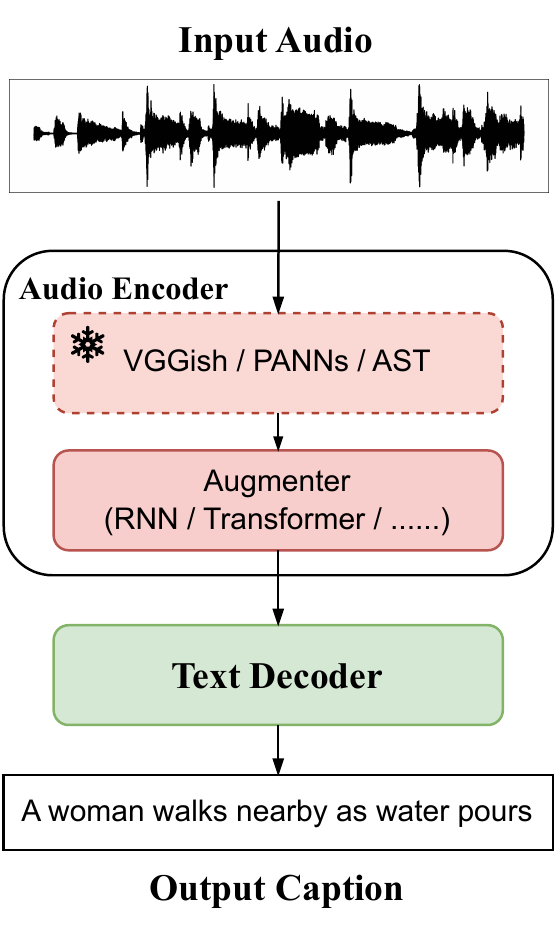}
    \caption{An illustration of AAC architectures leveraging a frozen pre-trained deep model by appending a trainable augmenter.}
    \label{fig:freezing_pretrain_augmenter}
\end{figure}

\paragraph{Audio Encoder Pre-training}
The exploration of audio encoder pre-training is very common in AAC.
To better exploit the limited AAC data, researchers first extract keywords from captions and pre-train the audio encoder by keyword estimation~\cite{chen2020audio,ye2021improving}.
Xu \etal~\cite{xu2021investigating} investigated the effects of different pre-training conditions and found that pre-training CNN on AudioSet brought the most significant improvement.
Many off-the-shelf models pre-trained on AudioSet, such as VGGish~\cite{hershey2017cnn}, PANNs~\cite{kong2020panns} and AST~\cite{gong2021ast}, are taken as the AAC encoder~\cite{han2021automated,won2021cau,kouzelis2022efficient}.
It should be noted that pre-trained deep models are often too large to be successfully adapted to small-scale AAC datasets.
A common practice is to freeze the pre-trained model and append a trainable augmenter (RNN, Transformer or graph attention~\cite{xiao2023graph}), as shown in \Cref{fig:freezing_pretrain_augmenter}.
The combination of frozen pre-trained deep models and a trainable augmenter has become the mainstream architecture to fully leverage the off-the-shelf models, achieving SOTA performance~\cite{wu2023beats}.

\paragraph{Text Decoder Pre-training}
The text decoder can be pre-trained by the language modelling task, but the adaptation of pre-trained language models is not as straightforward as the pre-trained audio encoders.
The pre-training by audio tagging fits the goal of the AAC audio encoder as the encoder also needs to recognize sound events for caption generation.
However, there is a discrepancy between the goal of a language model $p(w)$ and the AAC text decoder, a conditional language model $p(w|\mathcal{A})$.
The former aims to generate the subsequent text based on the preceding one while the latter's generation depends on an additional condition of the audio $\mathcal{A}$ in addition to the preceding text.
Therefore, despite the success of pre-trained language models in NLP, they had faced challenges in achieving satisfactory performance on AAC for a long time.
Koizumi \etal~\cite{koizumi2020audio} and Liu \etal~\cite{liu2022leveraging} incorporated GPT-2~\cite{radford2019language} and BERT~\cite{devlin2019bert} into the text decoding process.
These attempts did not bring significant improvement.
They even found that compact BERT with fewer parameters achieves better results.
Recently, with prompt learning proposed in NLP, the adaptation of pre-trained deep language models to AAC achieves much better performance.
Kim \etal~\cite{kim2023prefix} and Schauml\"{o}ffel \etal~\cite{schaumloeffel2023peacs} both adopted prefix tuning~\cite{li2021prefix} to leverage GPT-2.
Audio embeddings are mapped to a sequence of prefix vectors, which is taken as the context input to GPT-2 for caption generation. 
Kim \textit{et al.} showed that prefix tuning improves the model's generalizability to different caption styles while Schauml\"{o}ffel \textit{et al.} showed that the performance can be further improved when GPT-2 is also trained.
Unlike audio encoder pre-training which generally works well, the effectiveness of transferring pre-trained language models highly depends on the adaptation method. 


Efforts of encoder pre-training and decoder pre-training indicate that effectively adapting pre-trained models to the target AAC task poses a challenge.
One straightforward approach is \textbf{fine-tuning the pre-trained models on the target task}.
However, directly fine-tuning the text decoder often yields sub-optimal results, due to the difference between objectives of the pre-training task and the target task.
Another approach is to \textbf{transform the target task into the pre-training task}.
For the audio encoder, the pre-training task naturally resembles the target task, as both audio tagging and AAC require recognizing sound events from the input audio.
Appending a trainable augmenter to the frozen pre-trained model works well.
For the text decoder, prefix tuning successfully transforms the target task (conditional language modelling) into the pre-training task (language modelling) by mapping the condition (audio embeddings) into a prefix sequence.
Similar to the pre-training task, the decoder generates the next word depending on only the previous ones, without attending to the audio embeddings by cross-attention.
Therefore, prefix tuning achieves significantly better results than directly fine-tuning language models.

%% file: chapters/training_schemes.tex
\section{Training Schemes}
\label{sec:training_schemes}

Besides the standard encoder-decoder framework with the XE training loss, a plethora of works have proposed new training schemes.
Most variations on training schemes can be divided into these categories:
a) extra guidance; b) reinforcement learning; c) audio-text self-supervised learning; d) diverse or controllable captioning; e) other topics.
Works regarding these training schemes are discussed in the following part.
Finally, we give a summary of their effectiveness.

\subsection{Extra Guidance}
\label{subsec:extra_guidance}

\begin{figure}
    \centering
    \includegraphics[width=0.5\linewidth]{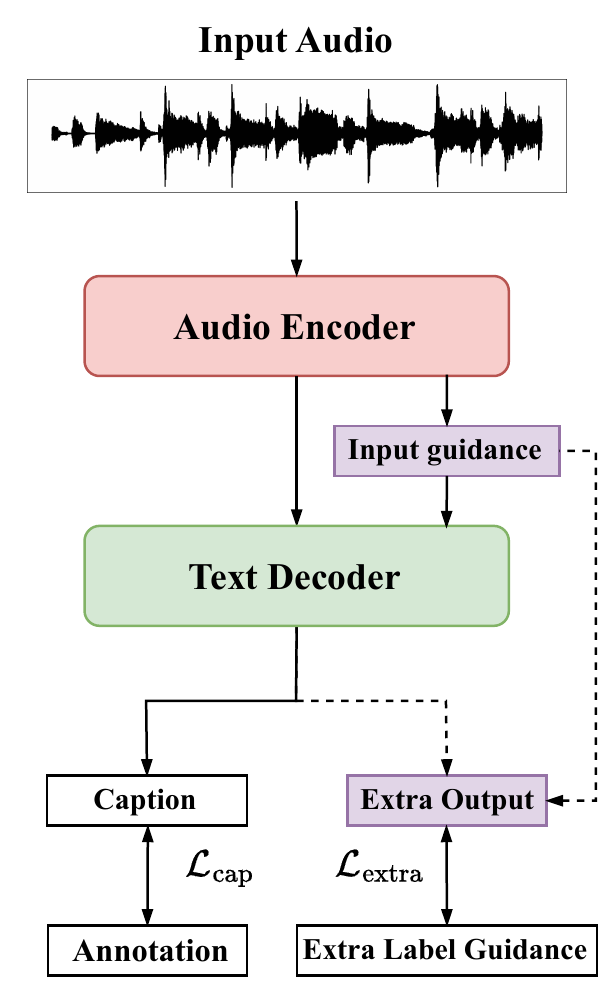}
    \caption{An illustration of using extra guidance for AAC. The AAC model can be guided through extra input or extra output. Dashes lines denote that the extra output can be obtained after either the audio encoder or the text decoder.}
    \label{fig:extra_guidance}
\end{figure}

To better generate accurate and detailed captions, researchers have made a lot of efforts to provide extra guidance besides the input audio and the annotated caption for AAC training.
The guidance works at the input or output of the AAC model, as shown in \Cref{fig:extra_guidance}.

In terms of extra input, several works aimed to provide additional information to guide the caption generation.
Keywords (or sound events) were the most used guidance since they indicate the main content of the caption.
The model first estimated description keywords (usually verbs and nouns) from the audio, and then generated captions based on estimated keywords or keyword probabilities~\cite{kim2019audiocaps,eren2020semantic,ye2021improving,kouzelis2022efficient}.
Some works used audio tagging results from pre-trained models, such as PANNs, as the keywords to ensure the reliability of the extra input guidance~\cite{gontier2021automated,eren2021dcase}.
To reduce the impact of keyword prediction errors, \cite{gontier2021automated} used keywords sampled from the output distribution as the extra input during training.
Other types of guidance were also explored, including the caption length~\cite{koizumi2020ntt}, PANNs embedding~\cite{narisetty2021leveraging} and similar captions~\cite{koizumi2020audio}.

Despite such exploration, the extra input guidance did not bring consistent improvement.
The estimation error of the input guidance significantly influences the final captioning performance~\cite{kim2019audiocaps}.
The little improvement may be attributed to the lack of extra information.
Such an approach works like a two-stage generation: first generating the main content, then the additional parts like conjunction words or sound event attributes to make sentences grammatically correct and enrich the content.
However, the extra input guidance is inferred from the audio so no extra information apart from the original input audio is incorporated.
Therefore, if the model providing extra guidance is trained on the same AAC dataset, the extra guidance offers little help.  
In contrast, when the extra guidance is from a model pre-trained on other datasets (e.g., AudioSet), it brings improvement since external knowledge is involved in caption generation.


In addition to extra input guidance, some works provide guidance from extra labels.
As shown in \Cref{fig:extra_guidance}, the extra output can be also the input guidance or be separately predicted besides the caption.
The model is trained in a multi-task scheme by combining the captioning loss $\mathcal{L}_\text{cap}$ and the extra loss $\mathcal{L}_\text{extra}$.
$\mathcal{L}_\text{extra}$ can be the keyword estimation loss~\cite{koizumi2020ntt,ye2022automated,cakir2020multi}, audio tagging loss~\cite{yan2023leveraging} and the sentence embedding loss~\cite{xu2021audio,mahfuz2023improving}.
$\mathcal{L}_\text{extra}$ based on audio-text matching is separately discussed in \Cref{subsec:audio_text_ssl}.
However, similar to extra input guidance, the extra label guidance does not work well under all scenarios. 
Takeuchi \etal~\cite{takeuchi2020effects} demonstrated that multi-task learning by training all models from scratch is not effective.
In comparison, the incorporation of pre-trained models~\cite{xu2021audio,mahfuz2023improving} to provide extra labels often yield significant improvements.

In summary, although the employment of extra guidance has been explored in many works, they do not show consistent advantages over the baseline without extra guidance.
By comparing successful approaches with others, it can be concluded that the incorporation of external knowledge (often from pre-trained models such as PANNs or BERT) is imperative for enhancing captioning performance.
If the extra input or label solely originates from the audio-caption data, it is likely that the guidance would offer limited assistance in caption generation.
For extra input guidance, the issue becomes even more pronounced since the prediction error of the extra input will accumulate and significantly impact the generated caption.

\subsection{Reinforcement Learning}
\label{subsec:reinforcement_learning}

\begin{figure}
    \centering
    \includegraphics[width=0.6\linewidth]{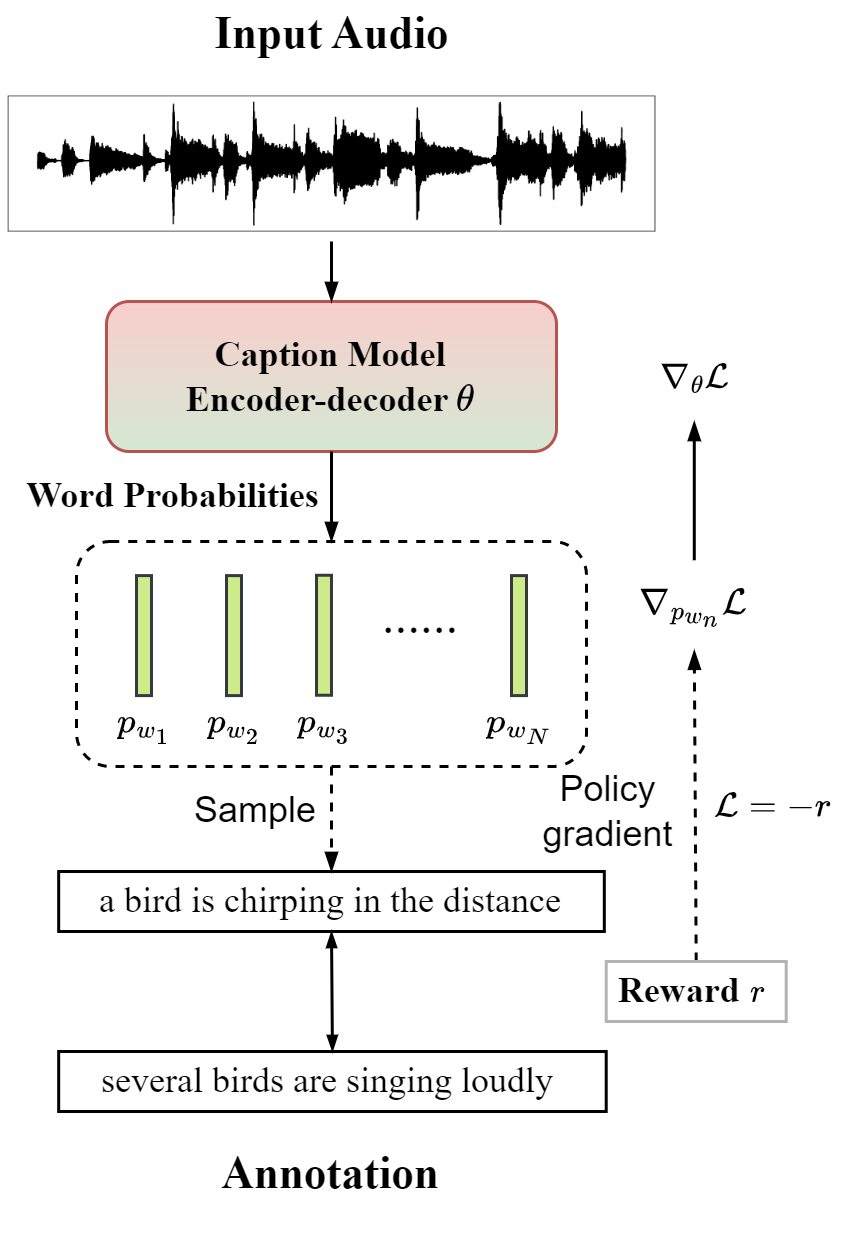}
    \caption{Illustration of audio captioning based on reinforcement learning. A caption is sampled given the word probabilities estimated by the model (with parameters $\theta$). Policy gradient is utilized to estimate the parameter gradients with the reward calculated given the annotation. Dashed lines denote sampling and gradient estimation while solid lines denote forward and backward calculation in neural networks.}
    \label{fig:rl_caption}
\end{figure}

Training by the standard word-level XE loss is often criticized for exposure bias~\cite{bengio2015scheduled} due to the teacher forcing strategy~\cite{williams1989learning}.
Reinforcement learning (RL) is adopted to alleviate this problem by directly optimizing the metric of the whole caption during training~\cite{xu2020crnn,mei2021encoder,ye2021improving}.
As shown in \Cref{fig:rl_caption}, a caption is sampled from word probabilities estimated by the captioning model.
The reward of the sampling process is calculated so that the gradients of the word probabilities can be estimated by policy gradient.
An effective variant of this approach, self-critical sequence training~\cite{rennie2017self} (SCST), is often employed in AAC, where a baseline was used to reduce the sampling variance.
The reward of the greedy decoded caption is chosen as the baseline:
\begin{equation}
    \nabla_{\theta}\mathcal{L} = -(r(s^s) - r(\hat{s}))\nabla_{\theta}\log p(s^s|\theta)
\end{equation}
where $s^s$ and $\hat{s}$ are sampled and greedy decoded captions, respectively.

RL has led to significant improvement in evaluation metrics.
In DCASE 2021 and 2022 challenges, most submissions used RL to fine-tune an XE-trained model.
However, as observed in visual captioning~\cite{li2019meta}, models trained by RL tend to generate captions with incorrect syntax though these captions could achieve high scores.
Such phenomenon is also observed in AAC~\cite{mei2021encoder,zhou2022can}.
For example, for a reference caption ``paper crackling with female speaking lightly in the background'', adding ``and in the'' after the candidate caption ``wind blowing lightly followed by a female speaking'' significantly improves scores since words like ``and'' and ``in'' frequently occur in annotations.
Adding these words makes the sentence grammatically erroneous while bringing nothing meaningful.
This means that models are actually over-fitting the evaluation metric and the metric is not able to accurately evaluate a caption's quality.
In the DCASE 2023 challenge, the grammar errors are penalized during evaluation and RL is not utilized in most submissions.
Current works using RL only improve unreliable metrics without enhancing the caption quality.
Optimizing a more reliable metric (see \Cref{subsec:eval_metrics}) using RL may be promising but the challenge lies in the training efficiency.
Since many newly-proposed metrics are based on pre-trained models instead of rules, the reward calculation may be too time-consuming to train.
Potential solutions to address this challenge need further exploration.

\begin{figure}[ht]
    \centering
    \includegraphics[width=\linewidth]{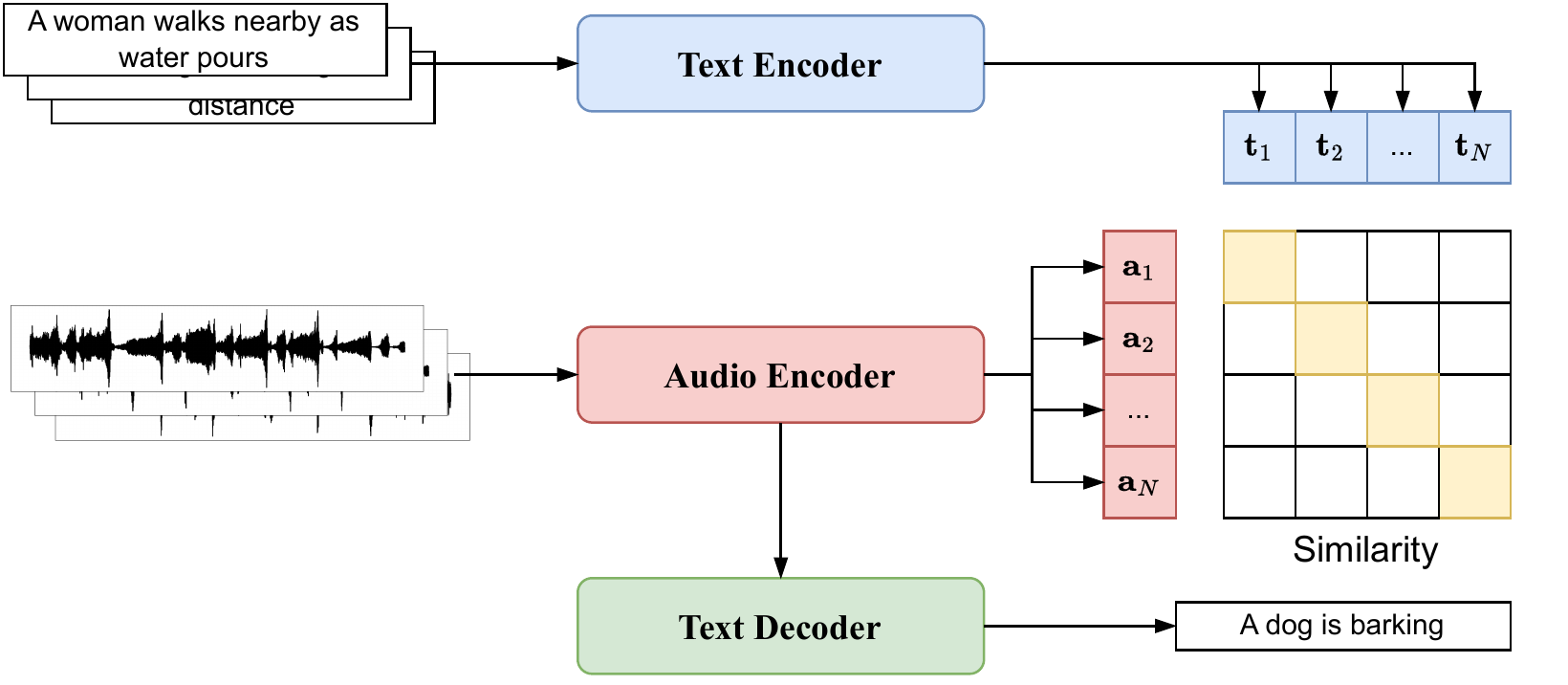}
    \caption{Illustration of CLIP-style audio-text self-supervised learning method. The audio encoder in the captioning model is trained by self-supervised contrastive learning. The objective is to bring paired audio and text embeddings closer while pushing unpaired embeddings apart.}
    \label{fig:audio_text_ssl}
\end{figure}

\subsection{Audio-text Self-supervised Learning}
\label{subsec:audio_text_ssl}
AAC datasets naturally provide the matching labels between audio and text: the audio and its corresponding annotation are paired while the audio is unpaired with other annotations.
The recent attention towards the utilization of audio-text self-supervised learning (SSL), which relies on such matching labels to facilitate audio captioning, has been increasing.
Although the matching loss is often used as $\mathcal{L}_\text{extra}$ in the multi-task learning scheme, we include them in this part because they all aim to learn the correspondence between audio and text.
Most of these works are based on contrastive learning.
Liu \etal~\cite{liu2021cl4ac} used the audio-text representation from the decoder output to predict whether an audio and a caption were paired.
Chen \etal~\cite{chen2022interactive} and Xu \etal~\cite{xu2022sjtu} took a CLIP-like approach where a separate text encoder extracted the text representation to calculate audio-text similarity, as \Cref{fig:audio_text_ssl} shows.
Zhang \etal~\cite{zhang2023actual} proposed a similar contrastive learning scheme to construct a proxy feature space where embeddings of captions belonging to the same audio clip are trained to be closer, while embeddings of captions from different audio clips are trained to be farther apart. 
During captioning training, the decoder's hidden embeddings are encouraged to be close to the caption's proxy embedding.
Apart from learning by discrimination, Koh \etal~\cite{koh2022automated} adopted another approach: learning by reconstruction.
The captioning model is trained to reconstruct the encoder embedding using the decoder embedding.
In this way, the similarity between encoder and decoder embeddings is maximized.

In summary, the employment of audio-text SSL, whether through discrimination or reconstruction, in AAC has consistently improved its performance.
Compared with using captions alone as training labels, audio-text SSL incorporates extra information (the matching/mismatch of audio-text pairs), thereby effectively enhancing the performance.

\subsection{Diverse / controllable Captioning}
\label{subsec:diverse_controllable_captioning}
Since humans describe an audio with different expressions, diverse captioning is investigated to make an AAC system more human-like.
Some of these works incorporate a control signal in the caption generation process and generate diverse captions by altering the control signal.

Ikawa \etal~\cite{ikawa2019neural} proposed a conditional AAC model where a ``specificity'' was input to the decoder to control the information contained in the generated caption.
A caption's specificity is inversely proportional to word frequencies so the caption with higher specificity contains more infrequent and informative words.
The model is trained to generate captions with a specificity value given the corresponding specificity input.
Xu \etal~\cite{xu2022diversity} took a similar architecture but the condition is calculated by a neural network during captioning training.
The neural network is trained to discriminate between human annotations and machine-generated captions.
They aimed to enhance the diversity of generated captions on the whole dataset while maintaining the description accuracy.
By altering the input specificity during inference, these two approaches could generate diverse captions with different detailedness.
In contrast to these two studies, the condition in Mei \etal~\cite{mei2022diverse} is a random Gaussian noise instead of a manually assigned value.
They employed a conditional generative adversarial network (cGAN) to generate diverse captions for the same input audio, which differs slightly from the objective in \cite{xu2022diversity}.

Although a few studies have worked on diverse and controllable captioning, there are some challenges.
\begin{itemize}
    \item The evaluation metric for diverse and controllable captioning is not unified. Conventional caption quality metrics (see \Cref{subsec:eval_metrics}) are also utilized in evaluating diverse captioning systems. However, the increase in diversity will almost definitely lead to the decrease in the conventional metrics. A commonly-used metric that can evaluate the caption quality and diversity simultaneously is lacking.
    \item Existing diverse captioning systems only concentrate on describing the same sound events through diverse linguistic expressions.
    Qualitative results~\cite{xu2022diversity,mei2022diverse} demonstrate that diversity is achieved by altering words or phrases, but different captions still describe the same sound events.
    Human-annotated captions are diverse in many aspects.
    Different annotators may describe a sound as semantically different but acoustically similar events (e.g., paper rustling or leaves rustling).
    They may also describe the audio from different levels: describing specific sound events (ball bouncing, whistling) or inferring the event (basketball match) and the environment (basketball stadium).
    Current approaches are still not able to achieve such diversity.
    More exploration on this topic should be made.
    We will further discuss it in \Cref{subsec:future_controllable_captioning}.
\end{itemize}

\begin{table*}[ht]
    \centering
    \caption{Comparison of AAC datasets. ``\#'' denotes the number. ``Avg'' and ``std'' denote average and standard deviation, respectively.}
    \begin{tabular}{c|c|c|c|c|c|c}
    \toprule
    \textbf{Dataset} & \textbf{Audio Source} & \textbf{Total duration / h} & \textbf{Clip Duration / s} & \textbf{\# Audios} & \textbf{\# Captions} & \textbf{Avg (std) caption length} \\ 
    \midrule
    AudioCaps & AudioSet & 139.58 & $\leq$ 10 & 51,308 & 57,188 & 8.79 (4.30)\\
    Clotho & Freesound & 37.06 & 15 $\sim$ 30 & 5,929 & 29,645 & 11.33 (2.78)\\
    MACS & TAU Urban Acoustic Scenes & 10.92 & 10 & 3,930 & 17,275 & 9.25 (3.89) \\
    \bottomrule
    \end{tabular}
    \label{tab:datasets}
\end{table*}

\subsection{Other Topics}
Apart from the training schemes described above, some works explore other schemes or objectives:
\begin{itemize}
    \item Loss modification: Cak{\i}r \etal~\cite{cakir2020multi} and Kothinti \etal~\cite{kothinti2022investigations} applied word frequency-based weights on the XE loss to address the vocabulary imbalance problem.
    Kothinti \etal~\cite{kothinti2022investigations} used the focal loss~\cite{lin2017focal} to assign class weights based on prediction errors.
    \item Curriculum learning: Koh \etal~\cite{koh2022curriculum} proposed a curriculum learning approach to AAC. During training, stopwords or infrequent words were partially removed from the target captions.
    The proportion of removed words was decreased gradually during training, since stopwords and infrequent words were considered hard to generate for AAC systems.
    \item Joint training with other tasks: Narisetty \etal~\cite{narisetty2022joint} joint trained speech recognition and AAC since both tasks aim to generate word sequences from an audio and utilize an encoder-decoder framework.
    \item Multi-modal AAC: Liu \etal~\cite{liu2022visually} incorporated visual information to facilitate AAC. They used an adaptive audio-visual attention module to leverage the possibly noisy visual clue.
    \item Detailed AAC: Xie \etal~\cite{xie2023enhance} proposed an AAC system based on a quantized temporal tag to enhance the temporal relationship details in captions.
    \item Continual learning: Berg \etal~\cite{berg2021continual} adapted the continual learning method to update a pre-trained AAC model on new data without forgetting the knowledge from the previous one.
\end{itemize}

\subsection{Training Schemes Summary}
\label{subsec:training_scheme_summary}
We hereby provide a brief summary of the above training schemes.
Extra guidance is not consistently effective.
Methods that leverage external knowledge from pre-trained models are generally effective, while the opposite holds true.
RL only improves the oriented metrics and does not seem to genuinely enhance the accuracy of the content in descriptions.
On the other hand, audio-text SSL consistently improves the captioning performance as it leverages the matching information between audio and text.
Diverse or controllable captioning has achieved some progress, but as a research direction, it still faces several challenges.

%% file: chapters/data_augmentation.tex
\section{Data Augmentation}
\label{sec:data_augmentation}

As stated in \Cref{subsec:visual_captioning}, the size of AAC datasets is small.
Therefore, researchers have been exploring data augmentation techniques in AAC to alleviate the data scarcity problem since the early stages.
Data augmentation techniques can be divided into audio augmentation, text augmentation and audio-text pair augmentation.

\subsection{Audio Augmentation}
Audio augmentation approaches do perturbations to the input audio, without altering the underlying audio content.
Such augmentation can be used in general audio classification tasks.
Specaugment~\cite{park2019specaugment} and specaugment++~\cite{wang2021specaugment++}, which randomly mask the audio at the input and the hidden space respectively, were successfully employed in AAC~\cite{chen2020audio,ye2021improving}.
Other common techniques like speed perturbation, pitch shift, adding noise and reverberation were also explored but did not improve the performance.
These perturbations possibly change the audio content so that the augmented audio does not match the caption.
For example, the caption does not contain ``noise'' but noise is added to the original audio. 
Ye \etal~\cite{ye2022featurecut} proposed an effective online data augmentation approach FeatureCut where acoustic features were partly discarded according to the decoder attention weights.
The Kullback-Leibler divergence (K-L divergence) between estimated word probabilities using the original and augmented data was minimized.

\subsection{Text Augmentation}
Similar to audio augmentation, text augmentation randomly perturbs the caption label.
Since a caption is composed of several discrete tokens (words), the caption is perturbed by altering words.
However, it is highly possible that the semantic meaning of a caption changes by random word substitution.
Koizumi \etal~\cite{koizumi2020ntt} proposed changing words with low TF-IDF to avoid replacing informative words.
Cho \etal~\cite{chang2023hyu} proposed synonyms substitution which substitutes random words with its synonym in WordNet~\cite{miller1995wordnet}.
These two approaches proved to work well in DCASE challenges.

The above methods focus on word-level substitution.
However, sentence-level paraphrasing has not been explored in augmenting AAC captions.
Paraphrasing sentences by back-translation or LLMs also seems promising but remains underexplored.

\subsection{Audio-Text Pair Augmentation}

\begin{figure}[htpb]
    \centering
    \includegraphics[width=\linewidth]{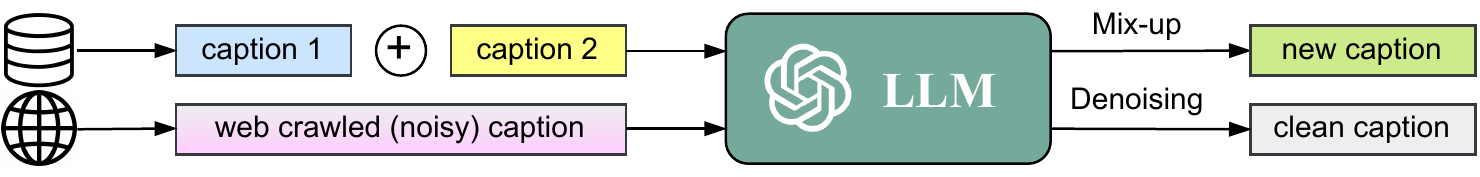}
    \caption{Large language model-based audio caption data augmentation.}
    \label{fig:llm_data_aug}
\end{figure}

Different from perturbation or rephrasing in a single modality, audio-text pair augmentation aims to create new cross-modal audio-text pairs.
As \Cref{fig:llm_data_aug} shows, there are two main types of audio-text pair augmentation techniques in existing works, where powerful LLMs can assist in both methods:
\begin{itemize}
    \item Mixing existing data. Mix-up is a common data augmentation technique to create a new data pair $(x, y)$ by linear interpolating existing pairs: $\lambda(x_1, y_1) + (1-\lambda)(x_2,y_2)$.
    However, the linear interpolation of text is not straightforward since it is a sequence of discrete tokens.
    Researchers proposed several solutions.
    Mix-up at the embedding space~\cite{koizumi2020ntt,kouzelis2022efficient} is utilized to mix different captions.
    Emmanouilidou \etal~\cite{kothinti2022investigations} concatenated captions using conjunction words.
    According to the conjunction words (``and'' or ``then''), the corresponding audio clips were mixed or concatenated.
    Wu \etal~\cite{wu2023beats} mixed two captions by ChatGPT~\cite{openai2021chatgpt}.
    With the help of ChatGPT, the mixed caption is grammatically correct while reflecting contents in the original two captions.
    \item Audio-text curation from the web. Noisy audio-text pairs naturally exist on the web. For example, user descriptions on Freesound~\cite{font2013freesound} can be used as noisy captions.
    Han \etal~\cite{han2021automated} designed heuristic rules to clean captions.
    Mei \etal~\cite{mei2023wavcaps} employed ChatGPT to convert Freesound descriptions into the style of captions in existing AAC datasets. 
\end{itemize}

Compared with single-modality augmentation, audio-text pair augmentation brings more significant improvement, especially approaches based on LLMs.
By incorporating LLMs into data augmentation, rich knowledge in massive textual data is leveraged to assist AAC, bringing new vitality to AAC research. 

%% file: chapters/datasets_evaluation.tex
\section{Datasets and Evaluation Metrics}

\label{sec:datasets_metrics}
In this section, several public AAC datasets are described in \Cref{subsec:datasets}.
Although there are AAC datasets on other languages~\cite{xu2021audio,ikawa2019neural} or specific domains~\cite{manco2021muscaps}, we focus on English, general domain AAC datasets since they are universally used as benchmarks.
We first briefly introduce their basic information as it is covered in \cite{mei2022automated} then analyze these datasets in detail by comparing them.

\Cref{subsec:eval_metrics} discusses AAC evaluation metrics.
These metrics include 1) conventional ones brought from machine translation and visual captioning; 2) recently-proposed ones designed specifically for AAC.
Finally, in \Cref{subsec:results} we list the evaluation results of some approaches on Clotho and AudioCaps.

\subsection{Datasets}
\label{subsec:datasets}
In this part, current datasets are introduced and compared. we first briefly describe current datasets and then compare them from the perspectives of the audio source and the caption style.
Statistics of these datasets are presented in \Cref{tab:datasets}.

\subsubsection{Dataset Description}
\paragraph*{AudioCaps}
AudioCaps~\cite{kim2019audiocaps} is the largest dataset for audio captioning in the wild.
The audios are a subset of AudioSet.
Annotators wrote descriptions with the AudioSet labels and video hints available. 
One caption is provided for the training subset while five annotations are provided for validation and test subsets.

\paragraph*{Clotho}
Clotho~\cite{drossos2020clotho} is the benchmark dataset used in DCASE challenges.
Audios are collected from Freesound~\cite{font2013freesound} platform, trimmed to 15 to 30 seconds and annotated in a three-step framework.
Each audio is provided with five annotations, ranging from eight to 20 words.

\paragraph*{MACS}
MACS~\cite{martin2021diversity} is a dataset originated from TAU Urban Acoustic Scenes 2019~\cite{mesaros2018multi}.
3,930 audios recorded in three scenes (airport, public square and park) are sampled and annotated.
Two to five annotations are provided for each audio.

\subsubsection{Dataset Comparison}

\paragraph*{\textbf{Audio source}}
As stated above, the three datasets stemmed from different sources.
They are all limited in sound event types: MACS contains only recordings in specific urban scenes so some sound events are not contained while AudioCaps and Clotho also exclude some sound events.
Developers of MACS found that Freesound samples often contain only the tagged sound event without background.
We also observe that a large proportion of Freesound samples contain only one sound event that lasts the whole clip without pause.
The operation to select segments with the highest energy in building Clotho~\cite{drossos2020clotho} exacerbates this phenomenon.
Therefore, audios in AudioCaps and MACS may contain richer temporal relationships between multiple sound events.
Besides the difference in audio diversity, the fact that AudioCaps is a subset of AudioSet, which provides sound event annotations, may encourage further research to make use of their connection.

\begin{figure*}[ht]
    \subfloat[AudioCaps]{
        \includegraphics[width=0.32\linewidth]{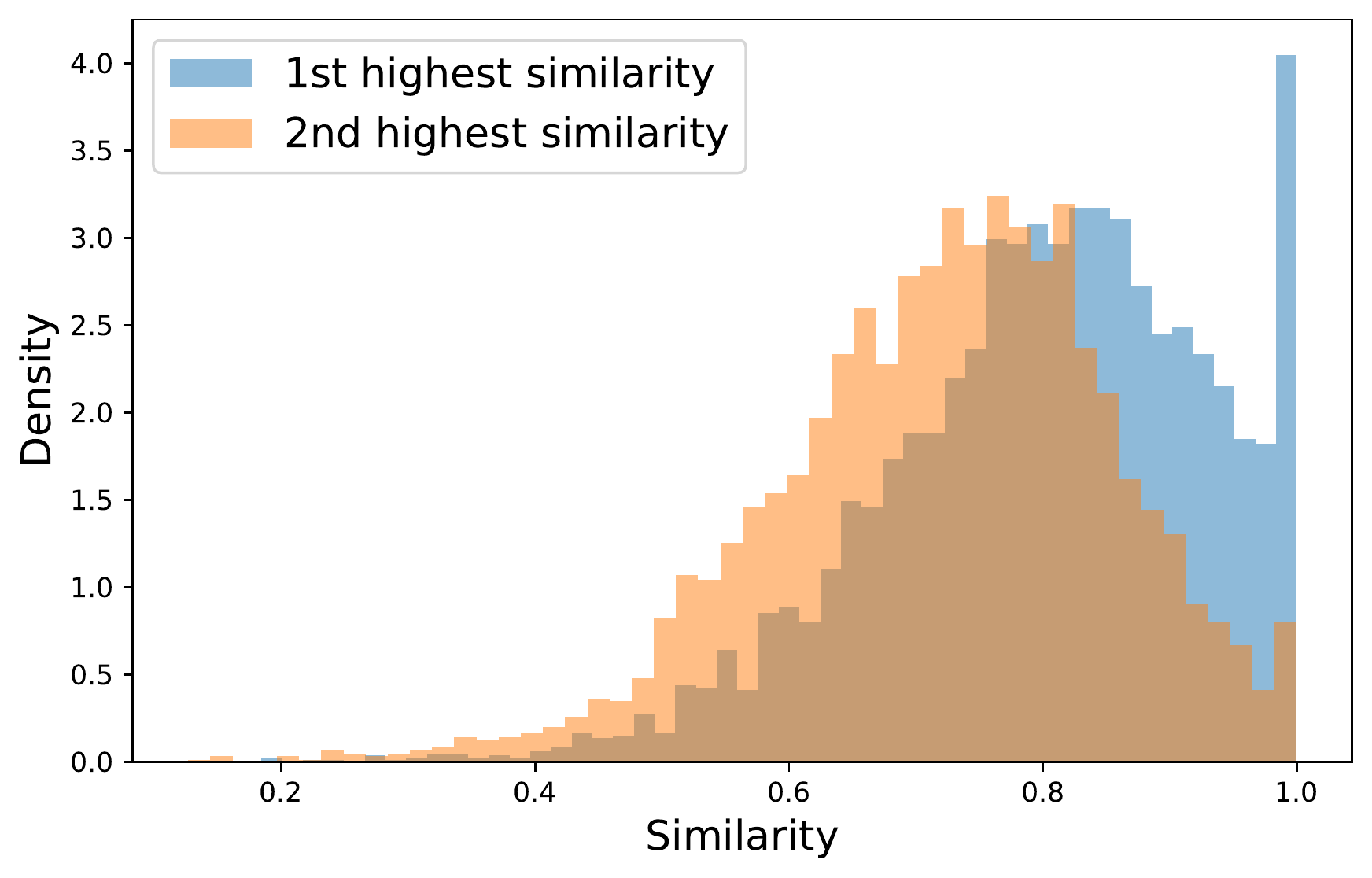}
    }
    \subfloat[Clotho]{
        \includegraphics[width=0.32\linewidth]{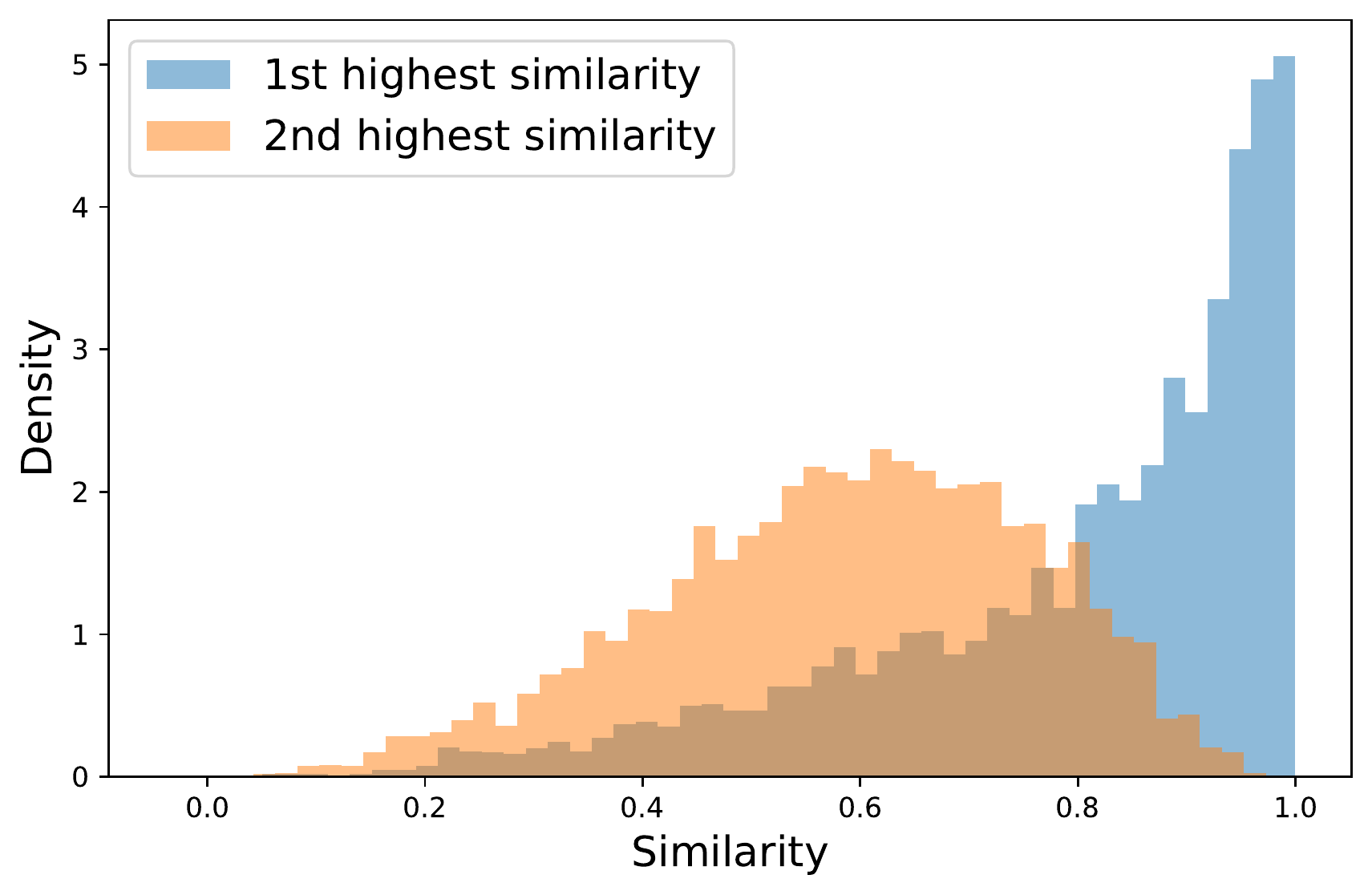}
    }
    \subfloat[MACS]{
        \includegraphics[width=0.32\linewidth]{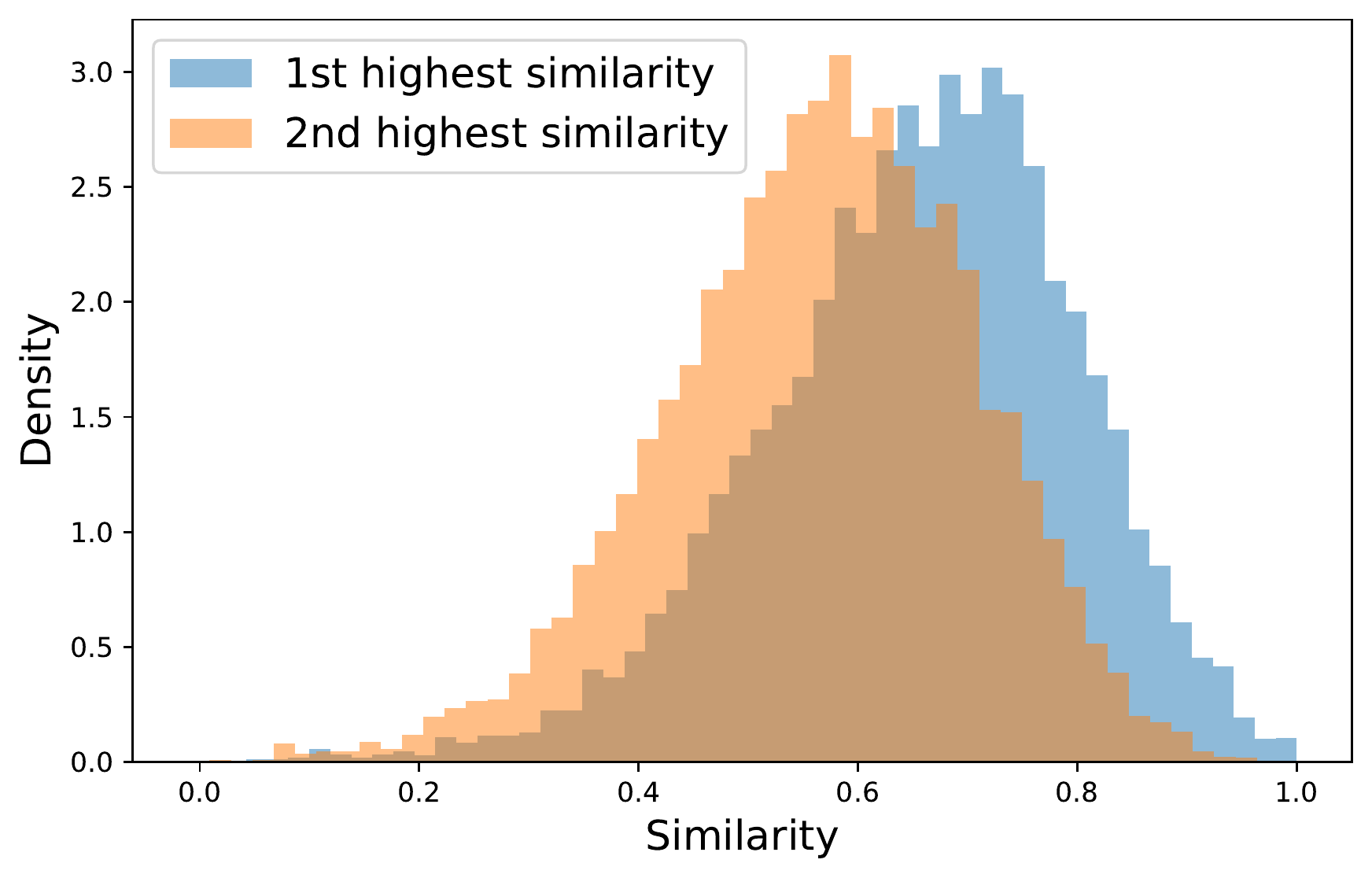}
    }
    \caption{The distribution of the highest and second-highest similarities between captions from the same audio for different datasets. There is a significant difference between the distribution of the highest similarity on Clotho and other distributions.}
    \label{fig:inter_cap_sim}
\end{figure*}

\paragraph*{\textbf{Caption style}}
Due to the difference in the annotating process, the captions of the three datasets differ in several aspects:
\begin{itemize}
    \item Visual information. The corresponding video of an audio is available in AudioCaps and annotators might use this hint to write captions. In contrast, the video is not available for Clotho and MACS. Therefore AudioCaps captions may contain visual information that is not available from the audio. Despite the inclusion of inaudible visual information into captions, the consensus among different annotators is enhanced, thus reducing the diversity of captions from different annotators (shown in the paragraph ``Caption diversity'' below). 
    \item Caption length variance. During Clotho annotation, the captions were restricted to contain eight to 20 words, making Clotho captions less varied in length (indicated by the standard deviation in \Cref{tab:datasets}). In contrast, AudioCaps and MACS both contain captions with only one word and captions with over 40 words.
    \item Description style. It is reported that AudioCaps captions are generally simpler, with shorter sentences and straightforward expressions~\cite{labbe2022my}.
    We observe that AudioCaps and MACS captions generally describe sound events faithfully.
    In contrast, the description style of Clotho captions significantly differs from them in two aspects: 1) it exhibits a more literary expression, employing sophisticated vocabulary and phrases. 2) the content extends beyond the sound events themselves, often encompassing imaginative descriptions of the environment. Inaudible information (e.g., at night) may also be included even if the video is unavailable.
    \item Caption diversity. Researchers observed that captions for the same audio in Clotho and MACS captions are semantically diverse~\cite{bhosale2023novel,zhang2023actual}.
    Since there is ambiguity in sound event types while annotators were asked to write informative captions, they tend to infer information, such as the environment, leading to description diversity.
    Quantitatively, Mei \etal~\cite{mei2022automated} calculated the consensus score to measure the agreement among different captions for the same audio.
    They concluded that MACS were more diverse than the other two.
    However, since the consensus score is calculated by comparing one caption with all other ones of the same audio, we assume it does not provide a complete picture of the caption diversity.
    We observe that the five captions for the same audio in Clotho can typically be divided into three groups.
    The first group contains two sentences.
    The second group contains another two sentences.
    The third group contains the rest.
    Captions from the same group were extremely similar with few words alternated but captions from different groups are very disparate.
    To give an illustration of this, we calculate the similarities (measured by SentenceBERT) between a caption and other captions for the same audio.
    We plot the distribution of the highest and the second-highest similarities for the same caption on the three datasets.
    The results in \Cref{fig:inter_cap_sim} indicate that there are much more extremely close caption pairs for the same audio in Clotho than AudioCaps and MACS.
    The high consensus scores on Clotho are caused by these close caption pairs.
    We assume the phenomenon is caused by the three-step annotation framework for Clotho.
    In the second step, annotators were asked to rephrase the captions from the first step.
    In the third step, captions from the previous two steps were put together and scored to select the top ones.
    Therefore, it was highly probable that the two captions were very similar since one was rephrased from another, and they were both selected in the third step.

    Given the distribution in \Cref{fig:inter_cap_sim} and our observation, we conclude the caption diversity of the three datasets as the following: 1) MACS is more diverse than AudioCaps. 2) The diversity of Clotho is similar to that of MACS (indicated by the peak similarity value of the second highest similarity distribution in the figure). However, the presence of extremely similar caption pairs of the same audio leads to Clotho's high consensus score in \cite{mei2022automated}.
    
\end{itemize}

\subsection{Evaluation Metrics}
\label{subsec:eval_metrics}
Automatically evaluating a generated caption is challenging since there are multiple plausible answers. 
Different references may focus on different sound events and may use different expressions. 
The free-form nature of natural language makes human evaluation remain the golden standard for natural language generation~\cite{celikyilmaz2020evaluation}.
Although automatic evaluation metrics are less reliable than human evaluation, they are still necessary since human evaluation is expensive and time-consuming.
We discuss these metrics here.
We first briefly introduce conventional metrics and then focus on discussing newly-proposed metrics.
They all give a score of a candidate caption based on several references.

\subsubsection{Conventional Metrics}
Conventional metrics are used in early DCASE challenges.
They are brought from the evaluation of other natural language generation tasks, including machine translation, text summarization and visual captioning.
BLEU, ROUGE and METEOR all measure n-gram overlaps between the candidate and references.
CIDEr and SPICE measure the semantic similarity between the candidate and references.
\begin{itemize}
    \item BLEU: Bilingual evaluation understudy (\textit{BLEU})~\cite{kishore2002bleu} was first proposed for machine translation evaluation. 
    It mainly measures \textbf{\textit{precision}}, i.e., the proportion of correctly generated n-grams in all generated n-grams.
    \item ROUGE: Recall-oriented understudy for gisting evaluation (\textit{ROUGE})~\cite{chin-yew2004rouge} was proposed to measure the quality of text summarization.
    ROUGE measures \textbf{\textit{recall}}, the proportion of correctly generated n-grams in the references.
    In AAC, one of its variants $\text{ROUGE}_\text{L}$ is used. 
    \item METEOR: Metric for evaluation of translation with explicit ordering (\textit{METEOR})~\cite{lavie2007meteor} is another metric for machine translation evaluation.
    It measures the F1 score between generated words and reference words with a penalty on wrong word ordering. 
    \item CIDEr: Consensus-based image description evaluation (\textit{CIDEr}) was proposed to evaluate visual captioning.
    It encodes a caption into an embedding using the term frequency-inverse document frequency (TF-IDF).
    Then the cosine similarity between the candidate's and references' embeddings is calculated as CIDEr.
    \item SPICE: Semantic propositional image caption evaluation (\textit{SPICE})~\cite{anderson2016spice} was also a visual captioning-oriented metric.
    It parses a caption into a scene graph~\cite{schuster2015generating} and extracts elements from the graph.
    SPICE is calculated as the F1 score between the elements extracted from the candidate and references.
    SPIDEr is the average of CIDEr and SPICE.
\end{itemize}

\subsubsection{AAC-oriented Metrics}
\label{subsubsec:aac_metrics}

Although the above conventional metrics are widely used, they are found to correlate badly with human judgments, hence not suitable for AAC evaluation.
Recently several metrics have been proposed to specifically evaluate AAC and enhance the reliability of the metric.
We present and discuss them in the following part.

\paragraph{FENSE}
Fluency ENhanced Sentence-bert Evaluation (FENSE)~\cite{zhou2022can} is similar to CIDEr in measuring the overall semantic similarity of two captions but it uses SentenceBERT~\cite{reimers2019sentence} instead of n-grams.
It calculates the cosine similarity of two captions' SentenceBERT embeddings.
Besides, an extra error detector was trained on synthetic data to penalize grammatically erroneous captions. 
FENSE is the combination of SentenceBERT similarity and grammar error penalization.
The comparison between FENSE and conventional metrics on their agreement with human judgments shows that FENSE is more reliable than conventional metrics.
Emmanouilidou \etal~\cite{kothinti2022investigations} also showed that FENSE was more robust to perturbation in captions than conventional metrics.
The grammar error penalization is a separate module and is also used in other metrics (SPICE+, SPIDEr-FL).

\paragraph{CB-score}
Similar to BLEU, content-based score (CB-score) is a \textbf{\textit{precision}}-type metric but measures the overlaps of sound events.
Captions are first mapped to sound events in AudioSet label sets by word matching considering synonyms and hypernyms in WordNet.
The consensus of different annotations is considered by assigning each sound event a relevance score based on its frequency in all annotations.
CB-score is calculated as the ratio between the candidate's relevance score (the sum of sound events' scores) and the sum of the top-$K$ relevance scores in references, where $K$ is the candidate's sound event number.
Since CB-score ignores the caption's lexical and grammar structure, it cannot be used alone but may be combined with other metrics to measure AAC.

\paragraph{s2vscore}
s2vscore~\cite{bhosale2023novel} is based on text-to-audio grounding (TAG)~\cite{xu2021text} to incorporate acoustic semantics into AAC evaluation.
The authors observed that multiple captions for the same audio may contain events that are semantically far apart but produce similar sounds, which aligns with our observation in \Cref{subsec:diverse_controllable_captioning}.
They developed a sound2vector (s2v) model based on TAG.
The s2v embeddings of acoustically similar sounds are close to each other.
Based on s2v, they transformed phrases in captions into s2v embeddings and calculated the F1 score between two captions.
The score is calculated in the same way as BERTScore~\cite{zhang2019bertscore} with the only difference that s2v embeddings replace BERT embeddings.
The results showed that s2vscore assigned higher scores to caption pairs from the same audio in Clotho compared with other metrics.

\paragraph{SPICE+}
SPICE+~\cite{gontier2023spice+} is a modification of SPICE proposed for AAC evaluation.
Two modifications are made to SPICE: 1) the original scene graph parser is replaced by a better deep language model-based parser with a refined set of linguistic rules to improve the parsing accuracy. 2) similar to the calculation of s2vscore, the original hard matching has been improved to a soft similarity score, so that different expressions describing the same sound event can get a non-zero score.
The correlation with human judgments shows that its combination with the fluency detector performs similarly to FENSE but SPICE+ provides better interpretability.

\paragraph{SPIDEr-max}
SPIDEr-max is different from all other metrics in that it evaluates several caption candidates.
It is an extension of SPIDEr by taking the maximum SPIDEr score of several candidates.
Since the authors observed that SPIDEr is highly sensitive to words but models may use synonyms in different captions, they proposed SPIDEr-max to take multiple caption candidates into account.
The same solution has also been applied in visual captioning~\cite{vijayakumar2018diverse} to evaluate diverse captioning.

\paragraph{SPIDEr-FL}
SPIDEr-FL is a combination of SPIDEr and the fluency detector proposed in FENSE.
It is the official metric used in the DCASE 2023 challenge.

In summary, these novel AAC-oriented metrics all take the characteristics of AAC into account, focusing on different aspects.
FENSE measures the overall semantic similarity of captions while also focusing on grammatical fluency.
However, only some pre-defined grammar error types are considered.
CB-score focuses on matching sound events but the extraction of sound events from captions contains inevitable errors.
SPICE+ uses a pre-trained deep language model to parse scene graphs and calculates a soft graph matching F1 score based on SentenceBERT.
s2vscore aims to assign high scores to acoustically similar while semantically disparate descriptions and calculates the F1 score in a manner similar to SPICE+.
Compared with other metrics using only the candidate and reference captions, it is the only metric involving the input audio in evaluation since the s2v embedding is calculated as the occurrence probability of a phrase in an audio.
SPIDEr-max is the only metric that measures a set of candidate captions.

\begin{table*}[htpb]
    \centering
    \caption{Performance of several approaches on Clotho and AudioCaps. TRM = Transformer, EG = Extra Guidance, MT = Multi-task Training, $\text{Pre}_\text{E}$ = Encoder Pre-training, $\text{Pre}_\text{D}$ = Decoder Pre-training, ATSSL = Audio-Text Self-Supervised Learning. ``Clotho v2*'' denotes that corresponding works do not follow the official train / validation split. They are all from DCASE challenges.
    }
    \begin{tabular}{c|cccccccc}
    \toprule
    \textbf{Dataset} & \textbf{Model Architecture} & \textbf{Training Schemes / Techniques} & $\textbf{BLEU}_\textbf{4}$ & $\textbf{ROUGE}_\textbf{L}$ & \textbf{METEOR} & \textbf{CIDEr} & \textbf{SPICE}\\
    \midrule
    \multirow{8}{*}{Clotho v1} & CNN + RNN~\cite{chen2020audio} & $\text{Pre}_\text{E}$  & 0.151 & 0.356 & 0.160 & 0.346 & 0.108 \\
    & RNN + RNN~\cite{cakir2020multi} & EG & 0.030 & 0.278 & 0.088 & 0.107 & 0.040 \\
    & CNN-Trm + Trm~\cite{koizumi2020transformer} & $\text{Pre}_\text{E}$, EG & 0.107 & 0.342 & 0.149 & 0.258 & 0.097 \\
    & CNN + RNN~\cite{xu2021investigating} & $\text{Pre}_\text{E}$ & 0.159 & 0.368 & 0.169 & 0.377 & 0.115 \\
    & CNN-Trm + Trm~\cite{koh2022automated} & $\text{Pre}_\text{E}$, ATSSL & 0.168 & 0.373 & 0.165 & 0.380 & 0.111 \\
    & CNN + Trm~\cite{chen2022interactive} & $\text{Pre}_\text{E}$, ATSSL & \textbf{0.169} & \textbf{0.379} & \textbf{0.171} & \textbf{0.407} & \textbf{0.119} \\
    & CNN + RNN~\cite{zhang2023actual} & $\text{Pre}_\text{E}$, ATSSL & 0.155 & 0.369 & \textbf{0.171} & 0.376 & 0.117 \\
    \midrule
    \multirow{9}{*}{Clotho v2*} & CNN + RNN~\cite{ye2021improving} & $\text{Pre}_\text{E}$, EG & 0.174 & 0.377 & 0.174 & 0.419 & 0.119 \\
    & CNN + RNN~\cite{xu2021sjtu} & $\text{Pre}_\text{E}$ & 0.155 & 0.374 & 0.174 & 0.399 & 0.119 \\
    & CNN + Trm~\cite{mei2021encoder} & $\text{Pre}_\text{E}$ & 0.174 & 0.379 & 0.171 & 0.426 & 0.124 \\ 
    & CNN-Trm + Trm~\cite{chen2021m2transformer} & $\text{Pre}_\text{E}$ & 0.153 & 0.366 & 0.168 & 0.409 & 0.120 \\
    & CNN + Trm~\cite{han2021automated} & EG, Data Augmentation & 0.182 & 0.389 & 0.177 & 0.474 & 0.135\\
    & CRNN + Trm~\cite{xu2022sjtu} & $\text{Pre}_\text{E}$, ATSSL & 0.166 & 0.378 & 0.179 & 0.421 & 0.127 \\
    & CNN + LSTM~\cite{ye2022automated} & $\text{Pre}_\text{E}$, EG &
    0.176 & 0.382 & 0.178 & 0.445 & 0.127 \\
    & CNN + Trm~\cite{mei2022dcase} & $\text{Pre}_\text{E}$, EG &
    0.167 & 0.382 & 0.175 & 0.409 & 0.123 \\    
    & CRNN + Trm~\cite{yan2023leveraging} & $\text{Pre}_\text{E}$, EG, ATSSL & - & - & 0.189 & 0.460 & 0.136 \\
    \midrule
    \multirow{15}{*}{Clotho v2}
    & Conformer + Trm~\cite{narisetty2021leveraging} & EG, Data Augmentation & 0.146 & 0.357 & 0.160 & 0.346 & 0.108\\
    & CNN + Trm~\cite{liu2021cl4ac} & $\text{Pre}_\text{E}$, ATSSL & 0.143 & 0.374 & 0.168 & 0.368 & 0.115 \\
    & Trm + Trm~\cite{kouzelis2022efficient} & $\text{Pre}_\text{E}$, EG, Data Augmentation & 0.173 & 0.387 & 0.178 & 0.457 & 0.134 \\
    & CNN + Trm~\cite{chen2022icnn} & $\text{Pre}_\text{E}$, EG & 0.180 & 0.383 & 0.175 & 0.443 & 0.123 \\
    & CNN + LocalAFT~\cite{xiao2022local} & $\text{Pre}_\text{E}$, Local Attention & 0.179 & 0.390 & 0.177 & 0.434 & 0.122 \\
    & CNN + RNN~\cite{zhang2023actual} & $\text{Pre}_\text{E}$, ATSSL & 0.161 & 0.375 & 0.176 & 0.409 & 0.121 \\
    & CNN + RNN~\cite{ye2022featurecut} & $\text{Pre}_\text{E}$, Data Augmentation & 0.179 & 0.389 & 0.176 & 0.436 & 0.122 \\
    & CNN-Trm + Trm~\cite{primus2022cpjku} & $\text{Pre}_\text{E}$, $\text{Pre}_\text{D}$, Data Augmentation & 0.158 & 0.376 & 0.181 & 0.440 & 0.128 \\
    & Trm-Trm + Trm~\cite{wu2023beats} & $\text{Pre}_\text{E}$, ATSSL, Data Augmentation & - & - & \textbf{0.193} & \textbf{0.506} & \textbf{0.146} \\
    & CNN + Trm~\cite{chang2023hyu} & $\text{Pre}_\text{E}$, Data Augmentation & 0.173 & \textbf{0.394} & 0.188 & 0.483 & 0.137 \\
    & CNN + Trm~\cite{labbe2023irit} & $\text{Pre}_\text{E}$, Data Augmentation & - & - & 0.192 & 0.485 & 0.139 \\
    & CNN + Trm~\cite{lee2023label} & $\text{Pre}_\text{E}$, Data Augmentation & 0.166 & 0.379 & 0.177 & 0.431 & 0.126 \\
    & Trm + Trm~\cite{schaumloeffel2023peacs} & $\text{Pre}_\text{E}$, $\text{Pre}_\text{D}$, Data Augmentation & - & 0.393 & 0.183 & 0.454 & 0.132 \\
    & CNN + Trm~\cite{kim2023prefix} & $\text{Pre}_\text{E}$, $\text{Pre}_\text{D}$ & 0.160 & 0.378 & 0.170 & 0.392 & 0.118 \\
    & CNN + Trm~\cite{mahfuz2023improving} & $\text{Pre}_\text{E}$, EG & 0.118 & 0.356 & 0.158 & 0.310 & 0.106 \\
    & CNN-Graph + Trm~\cite{xiao2023graph} & $\text{Pre}_\text{E}$, Graph Attention & \textbf{0.181} & 0.385 & 0.175 & 0.437 & 0.126 \\
    & Trm + Trm~\cite{mei2023wavcaps} & $\text{Pre}_\text{E}$, ATSSL, Data Augmentation & 0.168 & 0.383 & 0.184 & 0.462 & 0.133 \\ 
    \midrule
    \multirow{10}{*}{AudioCaps} & CRNN + RNN~\cite{kim2019audiocaps} & Pre, EG & 0.219 & 0.450 & 0.203 & 0.593 & 0.144\\
    & Trm-Trm~\cite{koizumi2020audio} & $\text{Pre}_\text{D}$, EG & 0.204 & 0.434 & 0.199 & 0.503 & 0.139 \\
    & CNN + GRU~\cite{xu2021investigating} & $\text{Pre}_\text{E}$ & 0.231 & 0.467 & 0.229 & 0.660 & 0.168 \\   
    & Trm + Trm~\cite{mei2021audio} & $\text{Pre}_\text{E}$ & 0.259 & 0.471 & 0.222 & 0.663 & 0.163 \\
    & Trm + Trm~\cite{gontier2021automated} & Pre, EG & 0.266 & 0.493 & 0.241 & 0.753 & 0.176 \\
    & CNN + Trm~\cite{liu2022leveraging} & $\text{Pre}_\text{D}$ & 0.251 & 0.480 & 0.232 & 0.667 & 0.172 \\ 
    & Trm + Trm~\cite{liu2022visually} & $\text{Pre}_\text{E}$, Visual Guidance & 0.281 & 0.494 & 0.237 & 0.711 & 0.172 \\
    & Trm + Trm~\cite{narisetty2022joint} & Joint Training with ASR & - & - & - & 0.632 & 0.171 \\
    & CNN + Trm~\cite{kim2023prefix} & $\text{Pre}_\text{E}, \text{Pre}_\text{D}$ & \textbf{0.309} & 0.503 & 0.240 & 0.733 & 0.177 \\
    & CNN + Trm~\cite{mahfuz2023improving} & $\text{Pre}_\text{E}$, EG & 0.208 & 0.455 & 0.218 & 0.574 & 0.166 \\
    & Trm + Trm~\cite{mei2023wavcaps} & $\text{Pre}_\text{E}$, ATSSL, Data Augmentation & 0.283 & \textbf{0.507} & \textbf{0.250} & \textbf{0.787} & \textbf{0.182} \\ 
    \bottomrule
    \end{tabular}
    \label{tab:results}
\end{table*}

These metrics are suitable for different scenarios.
FENSE, CB-score and SPICE+ can be used to measure a single candidate.
FENSE gives an overall quality evaluation.
CB-score and SPICE+ provide interpretability: why a candidate gets a low or high score? what erroneous content does this candidate include, or which content from references does it omit?
s2vscore and SPIDEr-max can evaluate diverse captioning systems since an ambiguous audio may be described as different sound events.
For a system generating diverse captions, s2vscore assigns high scores to a candidate as long as the sound events in the candidate sound similar to those in references.
SPIDEr-max reflects the quality of a set of generated diverse captions to some extent.
If one caption in the set is very similar to the given references, SPIDEr-max score will be high.
Not limited to SPIDEr, its maximum operation over several candidates can be used in other novel metrics for evaluating diverse captioning models.

\subsection{Performance Overview}
\label{subsec:results}

We summarize the experimental results of some works mentioned above in \Cref{tab:results}.
The corresponding model architectures and training schemes or techniques are also listed.
For works providing ensemble results, we present their single-model performance as it is more indicative of the effectiveness of methods.
We do not include results using RL due to its problem stated in \Cref{subsec:reinforcement_learning}.
It should be noted that there are no official train/validation splits for Clotho v1 and some challenge results are obtained by including official validation data into training, so we list them separately.
Since most works do not provide results using novel AAC-oriented metrics, we only present results based on conventional metrics.
Therefore the results are not reliable but we can still identify several trends. 
Results are consistent with our summary in \Cref{subsec:training_scheme_summary}.
Pre-training, audio-text SSL and data augmentation are generally effective approaches.
Different metrics focus on different aspects.
Methods that work best for one metric do not necessarily bring significant improvement in other metrics.

%% file: chapters/future_direction.tex
\section{Future Research Directions}

\label{sec:future_directions}
Although many works have been done towards AAC, there is still a large gap between machine-generated captions and human annotations regarding accuracy, diversity and usefulness.
With the aforementioned new metrics, there are several challenges that need to be addressed to foster the advancement of AAC research.
We discuss these challenges and potential research directions in this section.

\subsection{Re-evaluating New Metrics and Methods}

Despite several novel metrics proposed recently, the majority of current works still rely on conventional metrics to present results.
The latest DCASE challenge also adopts SPIDEr-FL, which is based on the conventional metric SPIDEr.
This practice limits the ability to accurately reflect the strengths and weaknesses of different methods in the current research.
Moreover, the benchmarks used in evaluating the reliability of metrics are not always consistent.
The human judgment data provided in FENSE contains a significant number of sentences generated by models with a low performance thus may be outdated.
It is necessary to supplement the human judgment data of comparing generated results from recent SOTA models.
We should compare the reliability of metrics such as SPICE+, FENSE, s2vscore on more up-to-date human judgment data, and re-evaluate current methods using these new metrics to obtain a more accurate picture of their performance.
Furthermore, we encourage future AAC research to report results using new metrics.

\subsection{Knowledge Transfer from Other Modalities}

Due to the scarcity of audio-text paired data and the abundance of data from other modalities such as text and vision, existing research has utilized knowledge from other modalities to assist AAC.
For instance, data augmentation approaches based on LLMs in \Cref{sec:data_augmentation} leverage rich knowledge from massive text corpora.
In addition to textual data, there is also a vast amount of image and image-text paired data available, which contains valuable knowledge.
Exploring how to effectively harness this knowledge to aid AAC and even broader audio-text tasks is an area worthy of further investigation.

\subsection{Model Compression and Acceleration}
Regardless of the architecture, current SOTA AAC models exhibit large model parameters and computational requirements, posing significant challenges for practical applications of AAC.
This challenge becomes even more pronounced when deploying AAC models on edge devices with limited computational resources.
Therefore, efforts in model compression, lightweight design, and inference acceleration are necessary.
Techniques such as knowledge distillation, quantization, and non-autoregressive models are worth investigating in this topic.
These advancements are crucial for the practical application of well-performing AAC models.

\subsection{Diverse / Controllable Captioning}

\label{subsec:future_controllable_captioning}
As stated in \Cref{subsec:visual_captioning}, the diversity of audio captions lies in the ambiguity of sounds.
In most Clotho samples, people describe the same audio differently because they cannot recognize the sound events clearly but have to imagine the sound source.
As a result, an AAC model is trained to generate disparate captions with exactly the same audio input.

We assume a controllable way of AAC may be more beneficial to leverage such diverse audio captioning data.
For example, the model may first recognize possible sound events and then describe (parts of) them with a control signal.
For an ambiguous sound, its possible sound events are acoustically similar but very disparate semantically.
For instance, when the model is confused about which of the two acoustically similar sound events (leaves rustling or paper rustling) appears in the audio, the caption styles can be controlled with an additional condition: 1) only describing sounds (rustling); 2) inferring sound sources (paper); 3) inferring the environment (in a room).
A controllable AAC model can also achieve diverse captioning.
By training the AAC model with condition signals corresponding to the reference captions, the challenge brought by the diversity of annotations may be alleviated.
The diversity of generated captions can also be extended from the variability in word and sentence expressions, which has been achieved to some extent, to the diversity in sound events and inferred environments.

The evaluation of such controllable AAC models is not the same as that of conventional AAC models.
The correspondence between audio and generated caption, the diversity and the controllability should all be evaluated.
Novel metrics like s2vscore and SPIDEr-max can be helpful in such a scenario.

\subsection{Detailed Captioning}

\label{subsec:future_detailed_captioning}
Current AAC models are found only to describe the main sound events with few details~\cite{xu2022diversity,mei2022diverse}.
It may be attributed to the characteristics of AAC data: people tend to only describe the prominent events in an audio without mentioning the event properties and relationships unless they are specifically asked to.
However, these details, especially event properties and temporal relationships between events, are vital to make a caption informative.
Without details, there is little difference between captions and audio tagging outputs, resulting in AAC's failure to provide effective assistance to users in human-computer interaction.
The inclusion of details in AAC is also crucial for real-world applications and warrants further investigation.
Xie \etal~\cite{xie2023enhance} has explored enhancing the temporal details in AAC.
Other details (see \Cref{sec:intro}) like low-level sound event attributes, induced human emotion, or even details like speech contents and music styles remain underexplored.
Given the limited presence of such details in current datasets, new audio captioning datasets that include descriptions of sound details are necessary.
In addition, knowledge from ASR, music classification and text summarization may be integrated with a standard AAC model to facilitate detailed AAC.

\subsection{Revisiting the Definition of AAC}
Although recent novel metrics that utilize pre-trained models and focus on sound events have achieved higher consistency with human judgment, they still remain insufficient for AAC evaluation.
While a good AAC model should undoubtedly produce accurate, diverse, and fluent outputs, the most crucial aspect is its utility for humans, as one of the purposes of AAC is to assist individuals with hearing impairments.
Therefore, it is important to first discuss what constitutes a useful caption for humans.
As stated above, we assume that a useful AAC system should be capable of describing details and even inferring environmental context.
Current datasets and evaluation metrics are not sufficient to develop such an AAC system.
Therefore, it is crucial to revisit the definition of AAC, which is the basis for developing new, useful and reliable datasets and metrics.
Only by defining what constitutes a useful caption can we develop an AAC system that is genuinely required in human-machine interaction.

%% file: chapters/conclusion.tex
\section{Conclusion}
\label{sec:conclusion}
In this paper, we make a comprehensive survey on AAC.
We present the characteristics of AAC from its definition and its comparison with other audio understanding and cross-modal text generation tasks.
We introduce the standard encoder-decoder architecture and discuss several pre-training practices.
Plenty of novel AAC training schemes are presented.
Data augmentation techniques, especially LLM-based methods, are introduced.
We give our summary and analysis on the effectiveness of these methods.
Datasets and evaluation metrics along with their characteristics and shortcomings are discussed.
Although remarkable progress has been achieved, the current AAC system outputs are far from accurate, detailed, diverse and useful to humans.
There are still several crucial challenges to be tackled.
We discuss some potential research directions to encourage novel works.
The survey is meant to provide a comprehensive introduction and insights about this task to new researchers.